\documentclass[iop]{emulateapj}

\def\lapp{\ifmmode\stackrel{<}{_{\sim}}\else$\stackrel{<}{_{\sim}}$\fi}
\def\gapp{\ifmmode\stackrel{>}{_{\sim}}\else$\stackrel{>}{_{\sim}}$\fi}

\newcommand{\frb}{FRB~121102}
\newcommand{\source}{FRB~121102}
\newcommand{\src}{\source}
\newcommand{\chandrasrc}{CXOU~J053156.7+330807}

\newcommand{\cxo}{\textit{Chandra}}
\newcommand{\chandra}{\textit{Chandra}}

\newcommand{\swift}{\textit{Swift}}

\newcommand{\dmunits}{\,pc\,cm$^{-3}$}
\newcommand{\micro}{\upmu}

\newcommand{\dnud}{\Delta\nu_{d}}
\newcommand{\pbj}{FRBCAT}

\usepackage{amsmath}
\usepackage{upgreek}
\usepackage{longtable}
\usepackage{lineno}
\usepackage[colorlinks,allcolors=blue]{hyperref}

\begin{document}

\title{The repeating Fast Radio Burst \src: Multi-wavelength observations and additional bursts}
\shorttitle{The repeating Fast Radio Burst \src}

\author{
P.~Scholz\altaffilmark{1},
L.~G.~Spitler\altaffilmark{2},
J.~W.~T. Hessels\altaffilmark{3,4}, 
S.~Chatterjee\altaffilmark{5},
J.~M.~Cordes\altaffilmark{5},
V.~M.~Kaspi\altaffilmark{1},
R.~S.~Wharton\altaffilmark{5},
C.~G.~Bassa\altaffilmark{3},
S.~Bogdanov\altaffilmark{7},
F.~Camilo\altaffilmark{7,8},
F.~Crawford\altaffilmark{9},
J.~Deneva\altaffilmark{10},
J.~van~Leeuwen\altaffilmark{3,4}, 
R.~Lynch\altaffilmark{11},
E.~C.~Madsen\altaffilmark{1},
M.~A.~McLaughlin\altaffilmark{12},
M.~Mickaliger\altaffilmark{6},
E.~Parent\altaffilmark{1},
C.~Patel\altaffilmark{1},
S.~M.~Ransom\altaffilmark{13},
A.~Seymour\altaffilmark{14},
I.~H.~Stairs\altaffilmark{15,1}, 
B.~W.~Stappers\altaffilmark{6},
\& S.~P.~Tendulkar\altaffilmark{1}
}

\altaffiltext{1}{Dept.~of Physics and McGill Space Institute, McGill Univ., Montreal, QC H3A 2T8, Canada; \href{mailto:pscholz@physics.mcgill.ca}{pscholz@physics.mcgill.ca}}
\altaffiltext{2}{Max-Planck-Institut f\"ur Radioastronomie, Auf dem H\"ugel 69, 53121 Bonn, Germany}
\altaffiltext{3}{ASTRON, the Netherlands Institute for Radio Astronomy, Postbus 2, 7990 AA Dwingeloo, The Netherlands; \href{mailto:J.W.T.Hessels@uva.nl}{J.W.T.Hessels@uva.nl}}
\altaffiltext{4}{Anton Pannekoek Institute for Astronomy, Univ. of Amsterdam, Science Park 904, 1098 XH Amsterdam, The Netherlands}
\altaffiltext{5}{Dept. of Astronomy and Cornell Center for Astrophysics and Planetary Science, Cornell Univ., Ithaca, NY 14853, USA} 
\altaffiltext{6}{Jodrell Bank Centre for Astrophysics, Univ.~of Manchester, Manchester, M13 9PL, UK} 
\altaffiltext{7}{Columbia Astrophysics Laboratory, Columbia Univ., New York, NY 10027, USA} 
\altaffiltext{8}{SKA South Africa, Pinelands, 7405, South Africa} 
\altaffiltext{9}{Dept.~of Physics and Astronomy, Franklin and Marshall College, Lancaster, PA 17604-3003, USA} 
\altaffiltext{10}{National Research Council, resident at the Naval Research Laboratory, 4555 Overlook Avenue SW, Washington DC 20375, USA} 
\altaffiltext{11}{National Radio Astronomy Observatory, PO Box 2, Green Bank, WV 24944, USA} 
\altaffiltext{12}{Dept.~of Physics and Astronomy, West Virginia Univ., Morgantown, WV 26506, USA} 
\altaffiltext{13}{National Radio Astronomy Observatory, Charlottesville, VA 22903, USA} 
\altaffiltext{14}{Arecibo Observatory, HC3 Box 53995, Arecibo, PR 00612, USA}
\altaffiltext{15}{Dept.~of Physics and Astronomy, Univ.~of British Columbia, Vancouver, BC V6T 1Z1, Canada} 

\begin{abstract}

We report on radio and X-ray observations of the only known repeating 
Fast Radio Burst (FRB) source, \frb.
We have detected six additional radio bursts from this source: five with the 
Green Bank Telescope at 2\,GHz, and one at 1.4\,GHz with the Arecibo Observatory
for a total of 17 bursts from this source. All have dispersion measures 
consistent with a single value ($\sim 559$\,pc\,cm$^{-3}$) that is three times 
the predicted maximum Galactic contribution.
The 2-GHz bursts have highly variable spectra like those at 1.4\,GHz,
indicating that the frequency structure seen across the individual 1.4 and 2-GHz 
bandpasses is part of a wideband process. X-ray observations of
the \frb\ field with the {\it Swift} and {\it Chandra} observatories show
at least one possible counterpart; however, the probability of chance superposition is high.
A radio imaging observation of the field with the Jansky Very Large Array
at 1.6\,GHz yields a 5$\sigma$ upper limit of 0.3\,mJy on any point-source 
continuum emission.  This upper limit, combined with archival
WISE 22-$\micro$m and IPHAS H$\alpha$ surveys, rules out the presence of an 
intervening Galactic \ion{H}{2} region.  
We update our estimate of the FRB detection rate in the PALFA survey to be 
$1.1^{+3.7}_{-1.0} \times 10^4$\,FRBs\,sky$^{-1}$\,day$^{-1}$ (95\% confidence)
for peak flux density at 1.4\,GHz above 300\,mJy.
We find that the intrinsic widths of the 12 \src\ bursts from Arecibo are, 
on average, significantly longer than the intrinsic widths of the 
13 single-component FRBs detected with the Parkes telescope.

\end{abstract}

\keywords{pulsars: general --- stars: neutron --- radio continuum: general --- X-rays: general}

\section{Introduction}

Fast Radio Bursts (FRBs) are an emerging class of astrophysical
transients whose physical origin is still a mystery.  They are
relatively bright (peak fluxes $\sim 0.5-1$\,Jy at 1.4\,GHz),
millisecond-duration radio bursts with high dispersion measures (DMs
$\gtrsim 300$\,pc\,cm$^{-3}$) that significantly exceed the maximum expected
line-of-sight contribution in the NE2001 model of Galactic electron density \citep{cl02}, and are thus thought to be
extragalactic in origin.  The distances implied by their DMs,
assuming that the excess dispersion is dominated by the intergalactic medium
(IGM) and a modest contribution from the host galaxy, place them at
cosmological redshifts \citep[e.g.,][]{tsb+13}.  Alternatively, if the majority
of the DM comes from near the source, they could be located in galaxies
at distances of tens to hundreds of megaparsecs \citep[e.g.,][]{mls+15}.

With the exception of one FRB detected with the 305-m Arecibo telescope \citep{sch+14} and one with
the 110-m Robert C. Byrd Green Bank Telescope \citep[GBT;][]{mls+15}, 
all of the 17 currently known FRBs have been detected using the 64-m Parkes 
radio telescope
\citep{lbm+07,tsb+13,bb14,rsj15,pbb+15,cpk+15,kjb+16}. 
While Arecibo and GBT provide significantly higher raw sensitivity 
($\sim10$ and $\sim3$ times greater than Parkes, respectively), 
the comparatively large field of view of the Parkes telescope, 
combined with the survey speed of its 13-beam receiver and
large amount of time dedicated to searching for pulsars and FRBs,
has proven to be a big advantage for blind FRB searches.
\citet{pbj+16}, 
hereafter FRBCAT, present an
online catalog of the known FRBs and their 
properties\footnote{\url{http://www.astronomy.swin.edu.au/pulsar/frbcat/}}.

The first FRB detected at a telescope other than Parkes was found in 
data from the PALFA pulsar and fast transient survey \citep{cfl+06,lbh+15}, which uses the 7-pixel
Arecibo L-Band Feed Array (ALFA) receiver system.
The burst, \frb, was discovered in a survey pointing towards the Galactic anti-center and in the plane: $l \sim 175^{\circ}$, $b \sim -0.2^{\circ}$ \citep{sch+14}.
The DM was measured to be $557\pm2$\dmunits, three times in excess
of the Galactic line-of-sight DM predicted by the NE2001 model \citep{cl02}.

\citet{ssh+16} performed follow-up observations with the Arecibo telescope
in 2013 December and 2015 May--June using a grid of ALFA pointings around the 
position of the original \src\ burst.
In three separate 2015 May--June observations,
10 additional bursts were detected at a DM and position
consistent with the original detection \citep{ssh+16}
-- though the new localization suggests that the discovery observation 
detected the source in a sidelobe of the receiver.

Thus far, no Parkes or GBT-detected FRB has been observed to repeat,
despite dozens of hours of follow-up observations in some cases \citep{pjk+15,mls+15}.  The
cosmological distances sometimes assumed for these events, along with their apparent non-repeatability,
has led to many theories of FRB origins that involve cataclysmic
events.  Examples include the merger of neutron stars or white dwarfs
\citep{kim13}, or the collapse of a fast-spinning and anomalously massive
neutron star into a black hole \citep{fr14}.   The discovery of a repeating FRB shows that, for 
at least a subset of the FRB population, the origin of 
such bursts cannot be from a cataclysmic event.
Rather, they must be due to a repeating phenomenon such 
as giant pulses from neutron stars \citep{cw16,pc15} or bursts from magnetars \citep{pp13}.  
The lack of observed repetition from the Parkes-discovered sources could, 
in principle, 
be due to the Parkes telescope's lower sensitivity compared to that of the 
Arecibo telescope.  If so, it is possible that all FRBs have a common physical 
origin, but that the observed population of bursts is strongly biased by 
limited sensitivity. Indeed, scaling the signal-to-noise ratios of the bursts 
in \citep{ssh+16} reveals that only the brightest burst would have been detected 
by Parkes.

In stark contrast to the origin implied by the repeating \frb, \citet{kjb+16} have recently claimed
the detection of a fading radio afterglow associated with the Parkes-discovered FRB~150418
at a redshift of 0.5.
Unlike \src, this discovery suggests that some FRBs may indeed originate from
cataclysmic events, as the merger of neutron stars is the preferred explanation.
However, others have challenged the
afterglow association, suggesting that it may instead be unrelated flaring from an active
galactic nucleus (AGN) or scintillation of a steady source \citep{wb16,vrm+16,aj16}.  
If the conclusion that the association is unlikely is supported by continued 
monitoring of the FRB~150418 field, then there is no need yet to postulate two 
separate types of FRB progenitors.

The detected bursts from \frb\ have several peculiar properties which 
undoubtedly provide important clues to their physical origin.
First, they appear to arrive clustered in time: of the
10 bursts presented by \citet{ssh+16}, six were detected
within a $\sim 10$-min period, despite having at total of $\sim4$\,hr of
on-source time in the 2013 December and 2015 May--June follow-up campaigns.
No underlying periodicity in the arrival time of the bursts was found.
The bursts also displayed unusual and highly variable spectral properties:
some bursts brighten significantly towards the highest observed frequencies, whereas others
become much brighter towards lower frequencies.  \citet{ssh+16} characterized
this behavior with power-law flux density models 
($S_{\nu} \propto \nu^{\alpha}$, where $S_{\nu}$ is the flux density at frequency $\nu$) where the
observed spectral index varied between $\alpha \sim -10$ to $+14$.
Even more peculiar is that at least two of the bursts are poorly 
described by a power-law model and appear to have spectra 
that peak within the 322\,MHz-wide band of ALFA.
Lastly, the detected bursts show no obvious signs of scintillation or scattering, 
both of which could provide important insights into the distance and source environment.

In this paper, we present further follow-up observations of
\src\ and the surrounding field using the \swift\ and \chandra\ X-ray
telescopes, the Karl G. Jansky Very Large Array (VLA), and Arecibo, Green Bank, 
Lovell and Effelsberg radio telescopes. 
In Section \ref{sec:obs} we describe 
the observations taken and the resulting datasets. In Section \ref{sec:bursts} we outline our analysis of bursts detected
in Arecibo and GBT observations.
We present images of the field around \src\ from our VLA, \swift\ and \chandra\ observations, as well as archival
optical and high-energy observations in Section \ref{sec:multi}.
We revisit the question of whether the source could 
be Galactic in Section \ref{sec:hii}.
In Section \ref{sec:rate}, we present an updated FRB rate from the PALFA survey.
In Section \ref{sec:disc} we discuss the implications of
a repeating FRB as well as the properties of our bursts 
in the context of other FRB detections.  

\begin{figure*}[t]
\includegraphics[width=\textwidth]{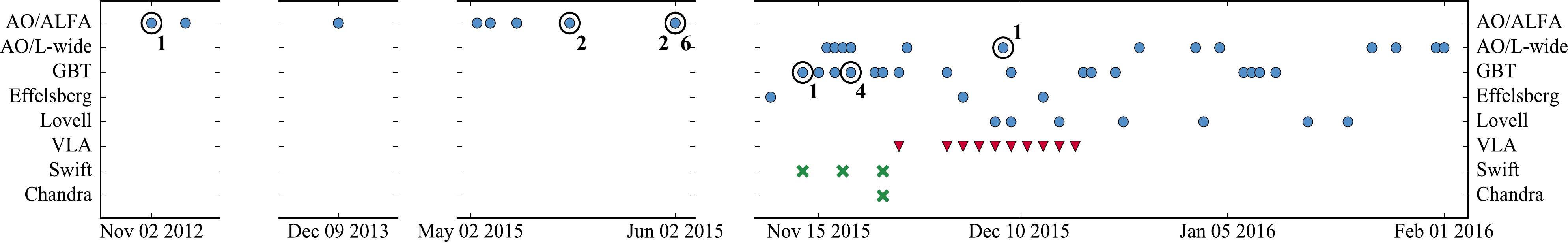}
\figcaption{
Timeline of radio and X-ray observations of \src\ from discovery in 
2012 November to 2016 January. Each row represents a set of observations from
a given telescope (and receiver in the case of Arecibo observations). 
Blue circles represent single-dish
radio observations, red triangles denote radio interferometer observations
and green crosses are X-ray observations. 
Observations with bursts are encircled and marked with the numbers of bursts
discovered. Note that bursts were discovered on 2015 June 2 in two separate
observations.
\label{fig:timeline}
}
\end{figure*}

\begin{deluxetable*}{ccccccccc} 
\tabletypesize{\normalsize} 
\tablecolumns{8} 
\tablewidth{0pt} 
\tablecaption{ Summary of Radio Telescope Observations \label{tab:scopes}} 
\tablehead{ 
\colhead{Telescope} & \colhead{Receiver} & \colhead{Gain}    & \colhead{T$_\mathrm{sys}$} & \colhead{Bandwidth} & \colhead{Central Frequency} & \colhead{Beam FWHM} & \colhead{Total Time} & Sensitivity\tablenotemark{i}\\ 
\colhead{}          & \colhead{}         & \colhead{(K/Jy)}  & \colhead{(K)}              & \colhead{(MHz)}     & \colhead{(MHz)}             & \colhead{(\arcmin)} & \colhead{on-source (hr)} & (Jy)   }
\startdata 
Arecibo             & ALFA\tablenotemark{a}    & 8.5 & 30 & 322  & 1375 & 3.4 & 4.4 & 0.02 \\
Arecibo             & L-Wide\tablenotemark{b}  & 10  & 30 & 700\tablenotemark{h}  & 1430 & 3.1 & 18.3 & 0.01 \\
GBT                 & S-band\tablenotemark{c}  & 2.0 & 20 & 800\tablenotemark{h}  & 2000 & 5.8 & 15.3 & 0.04 \\
GBT                 & 820~MHz\tablenotemark{c} & 2.0 & 25 & 200  & 820 & 15 & 7.4 & 0.09 \\
Effelsberg          & S60mm\tablenotemark{d}   & 1.55 & 27 & 500  & 4850 & 2.4 & 9.7 & 0.08 \\
Lovell              & L-band                   & 0.9 & 27 & 400  & 1532 & 12 & 14.0 & 0.15 \\
Jansky VLA          & L-band\tablenotemark{e}  & 2.15 & 35 & $2\times128$ & 1436, 1800 & 0.5--0.75\tablenotemark{g} & 10.0 & 0.1 \\ 
Parkes              & MB20\tablenotemark{f}    & 0.9 & 27 & 400 & 1380 & 14 & - & 0.15  
\enddata 
\tablenotetext{a}{\url{http://www.naic.edu/alfa/gen_info/info_obs.shtml}}
\tablenotetext{b}{\url{http://www.naic.edu/~astro/RXstatus/Lwide/Lwide.shtml}}
\tablenotetext{c}{\url{https://science.nrao.edu/facilities/gbt/proposing/GBTpg.pdf}}
\tablenotetext{d}{\url{https://eff100mwiki.mpifr-bonn.mpg.de/doku.php?id=information_for_astronomers:rx:s60mm}}
\tablenotetext{e}{\url{https://science.nrao.edu/facilities/vla/docs/manuals/oss2015B}}
\tablenotetext{f}{\url{https://www.parkes.atnf.csiro.au/cgi-bin/public_wiki/wiki.pl?MB20}}
\tablenotetext{g}{For synthesized beam}
\tablenotetext{h}{RFI filters reduce the usable bandwidth to $\sim600$\,MHz}
\tablenotetext{i}{Limiting peak flux sensitivity for a pulse width of 1.3\,ms and a S/N threshold of five.}
\end{deluxetable*}

\section{Observations}
\label{sec:obs}

Following the Arecibo discovery of repeat bursts from \src\ 
\citep{ssh+16},
we performed additional follow-up observations using a variety of telescopes. 
Unless otherwise noted, each telescope was pointed at the average position 
of the 2015 May--June detections from
\citet{ssh+16}: i.e. RA=$05^h31^m58^s$ and Dec=$+33\arcdeg08\arcmin04\arcsec$.
\citet{ssh+16} quote a conservative uncertainty region of 6\arcmin\ in diameter, 
approximately twice the full-width at half maximum (FWHM) of an ALFA beam at 1.4\,GHz.
Table \ref{tab:scopes} summarizes the observing set-ups used and Table \ref{tab:obs} 
lists all radio observations of \src, including the ALFA discovery and 
follow-up observations presented by \citet{ssh+16}. 
Figure \ref{fig:timeline} shows a timeline of the radio and X-ray observations
that we performed.

\subsection{Arecibo Telescope}
\label{sec:aoobs}

We observed \frb\ with the 305-m William E. Gordon
Telescope at the Arecibo Observatory (AO) using the
single-pixel L-Wide receiver ($1150-1730$\,MHz) and the Puerto-Rican
Ultimate Pulsar Processing Instrument (PUPPI) backend. We used PUPPI's coherent
filterbank mode, in which each of seven 100-MHz bands are sampled with
10.24-$\micro$s time resolution and 64 spectral channels.  Each of the
spectral channels was coherently dedispersed at DM =
557.0\,pc\,cm$^{-3}$.  As such, the recorded signals do not suffer significantly
from intra-channel dispersive smearing 
\citep[unlike the ALFA observations presented in][which use the Mock 
spectrometers\footnote{\url{http://www.naic.edu/∼astro/mock.shtml}}
 and incoherent dedispersion]{ssh+16}.  
Observing sessions ranged
from roughly $1-2$\,hr in length (\frb\ is only visible to Arecibo for 
$\sim 2$\,hr per transit), and typically we observed a 2-min
scan on a test pulsar, followed by a single long pointing at the best
known position of \frb.  
On a few occasions we observed
alternate pointing positions consistent with the 2015 May/June detection beams
reported by \citet{ssh+16} -- i.e. positions offset by a few arcminutes with 
respect to the average position quoted above.  
All scans were preceded by a 60-s calibration observation of a pulsed noise diode.  
The details of each session are given in Table \ref{tab:obs}.

\subsection{Green Bank Telescope}
\label{sec:gbtobs}

We observed \frb\ with the 110-m Robert C. Byrd Green Bank Telescope (GBT) 
using the Green Bank Ultimate Pulsar
Processing Instrument (GUPPI) backend and the 820-MHz and 2-GHz
receivers.  The 820-MHz observations used a
200-MHz bandwidth and recorded spectra every 20.48\,$\micro$s.  
The 2-GHz observations have 800\,MHz of nominal
bandwidth (RFI filters reduce the usable bandwidth to about 600\,MHz)
and the spectra were recorded every 10.24\,$\micro$s.  Note that the GBT beam has
a FWHM of $\sim 6\arcmin$ at 2\,GHz and thus 
comfortably encompasses the conservative positional uncertainty of \frb\
despite the higher observing frequency.
In each case, the data were
coherently dedispersed at the nominal DM of \frb\ and 512 spectral
channels were recorded with full Stokes parameters.  During a single
observing session, we observed \frb\ typically for 50\,min using
each receiver.  We also observed a pulsed noise diode at the start
of each scan for use in absolute flux calibration.  A detailed
description of each session is given in Table \ref{tab:obs}.


\subsection{Lovell Telescope}
\label{sec:jbobs}

We observed the position of \frb\ with the Lovell Telescope at the 
Jodrell Bank Observatory on seven separate epochs (see Table \ref{tab:obs})
for a total of 14\,hr.
Spectra were recorded with a total bandwidth of 400\,MHz over 800 channels
with a center frequency of 1532\,MHz, at a sampling time of 256\,$\micro$s. 
Unlike for the Arecibo and GBT observations, the spectral channels were not coherently dedispersed.
The data were RFI-filtered using a median absolute deviation 
algorithm before applying a channel mask, which results in typically
$\sim20$\% of the band being removed.

\subsection{Effelsberg Telescope}
\label{sec:effobs}

We observed the position of \frb\ with the Effelsberg 100-m Radio Telescope 
using the S60mm receiver, which covers the frequency range 4600--5100\,MHz, 
and the Pulsar Fast-Fourier-Transform Spectrometer (PFFTS) search backend. 
The FWHM of the Effelsberg telescope at 5\,GHz is 2.4\arcmin.  As such, the high-frequency
Effelsberg observations covered only the central region of the positional error box given in \citet{ssh+16}.
The PFFTS spectrometers generate total intensity spectra with a frequency 
resolution of 3.90625\,MHz and time resolution of $65.536\,\micro$s. 

\begin{table}
\begin{longtable}{ccccc}
\caption{Details of Radio Telescope Observations \label{tab:obs}} \\
\hline\hline
Date       & Start Time & Telescope/ & Obs. Length,  & No.\\ 
           & (UTC)      & Receiver   & $t_{obs}$ (s) & Bursts \\
\hline
\endfirsthead
\caption{(continued)}\\
\hline\hline 
Date       & Start Time & Telescope/ & Obs. Length,  & No.\\ 
           & (UTC)      & Receiver   & $t_{obs}$ (s) & Bursts \\
\hline
\endhead
2012-11-02 & 06:38:13 & AO/ALFA & 181  & {\bf1} \\
2012-11-04 & 06:28:43 & AO/ALFA & 181  & 0 \\
2013-12-09 & 04:09:52 & AO/ALFA & 2702 & 0 \\
2013-12-09 & 05:14:32 & AO/ALFA & 1830 & 0 \\
2013-12-09 & 04:55:19 & AO/ALFA & 970  & 0 \\
2015-05-03 & 18:55:48 & AO/ALFA & 1502 & 0 \\
2015-05-05 & 18:29:07 & AO/ALFA & 1002 & 0 \\
2015-05-05 & 19:39:15 & AO/ALFA & 1002 & 0 \\
2015-05-09 & 18:10:48 & AO/ALFA & 1002 & 0 \\
2015-05-09 & 19:20:56 & AO/ALFA & 1002 & 0 \\
2015-05-09 & 19:38:12 & AO/ALFA & 425  & 0 \\
2015-05-17 & 17:45:38 & AO/ALFA & 1002 & {\bf2} \\
2015-05-17 & 18:58:07 & AO/ALFA & 1002 & 0 \\
2015-06-02 & 16:38:47 & AO/ALFA & 1002 & {\bf2} \\
2015-06-02 & 17:48:52 & AO/ALFA & 1002 & {\bf6} \\
2015-06-02 & 18:09:18 & AO/ALFA & 300  & 0 \\
2015-11-09 & 22:36:47 & Effelsberg   & 9894 & 0 \\
2015-11-13 & 06:38:51 & GBT/820\,MHz & 3000 & 0\\
2015-11-16 & 05:24:09 & AO/L-wide    & 5753 & 0 \\
2015-11-13 & 07:42:09 & GBT/S-band   & 3000 & {\bf1} \\
2015-11-15 & 02:55:50 & GBT/S-band   & 3000 & 0 \\
2015-11-15 & 03:57:08 & GBT/820\,MHz & 3000 & 0\\
2015-11-17 & 03:24:33 & GBT/S-band   & 3000 &  0 \\
2015-11-17 & 04:34:40 & GBT/820\,MHz & 3000 & 0\\
2015-11-17 & 05:21:37 & AO/L-wide    & 6747 & 0 \\
2015-11-18 & 05:23:14 & AO/L-wide    & 6421 & 0 \\
2015-11-19 & 05:27:12 & AO/L-wide    & 3000 & 0 \\
2015-11-19 & 06:19:36 & AO/L-wide    & 1300 & 0 \\
2015-11-19 & 06:43:46 & AO/L-wide    & 971  & 0 \\
2015-11-19 & 10:14:57 & GBT/S-band   & 3000 & {\bf4} \\
2015-11-19 & 11:16:32 & GBT/820\,MHz & 1653 & 0\\
2015-11-22 & 08:45:23 & GBT/S-band   & 3000 & 0 \\
2015-11-22 & 09:47:15 & GBT/820\,MHz & 2543 & 0\\
2015-11-23 & 11:42:40 & GBT/S-band   & 5357 & 0 \\
2015-11-25 & 03:25:29 & VLA          & 3585 & - \\
2015-11-25 & 10:48:05 & GBT/S-band   & 5264 & 0 \\
2015-11-26 & 05:18:03 & AO/L-wide    & 2705 & 0 \\
2015-12-01 & 05:31:31 & VLA          & 3590 & - \\
2015-12-01 & 07:46:17 & GBT/S-band   & 6480 & 0 \\
2015-12-03 & 02:53:57 & VLA          & 3589  & - \\
2015-12-03 & 03:42:14 & Effelsberg   & 10800 & 0 \\
2015-12-05 & 04:45:44 & VLA          & 3589  & - \\
2015-12-07 & 04:37:50 & VLA          & 3589  & - \\
2015-12-07 & 21:36:17 & Lovell       & 7229 & 0 \\
2015-12-08 & 04:43:24 & AO/L-wide    & 3625 & {\bf1} \\
2015-12-09 & 00:01:52 & Lovell       & 7209 & 0 \\
2015-12-09 & 08:11:46 & GBT/S-band   & 3141 & 0 \\
2015-12-09 & 09:29:25 & VLA          & 3590 & - \\
2015-12-09 & 09:40:28 & GBT/820\,MHz & 3106 & 0\\
2015-12-11 & 09:22:27 & VLA          & 3590 & - \\
2015-12-13 & 00:38:29 & Effelsberg   & 14400 & 0 \\
2015-12-13 & 09:13:07 & VLA          & 3589  & - \\
2015-12-15 & 04:01:29 & Lovell       & 7141  & 0 \\
2015-12-15 & 09:06:16 & VLA          & 3590  & - \\
2015-12-17 & 08:57:51 & VLA          & 3589  & - \\
2015-12-18 & 08:55:38 & GBT/S-band   & 3600  & 0 \\
2015-12-18 & 10:08:41 & GBT/820\,MHz & 2216  & 0\\
2015-12-19 & 00:05:44 & GBT/S-band   & 3600  & 0 \\
2015-12-19 & 01:19:43 & GBT/820\,MHz & 1510  & 0\\
2015-12-22 & 00:11:13 & GBT/S-band   & 830   & 0 \\
2015-12-22 & 00:28:24 & GBT/S-band   & 2024  & 0 \\
2015-12-22 & 01:15:13 & GBT/820\,MHz & 2688  & 0\\
2015-12-23 & 04:55:57 & Lovell       & 7225 & 0 \\
2015-12-25 & 04:19:06 & AO/L-wide    & 1534  & 0 \\
2016-01-01 & 02:23:17 & AO/L-wide    & 5858  & 0 \\
2016-01-01 & 04:15:37 & AO/L-wide    & 46    & 0 \\
2016-01-02 & 01:16:52 & Lovell       & 7207 & 0 \\
2016-01-04 & 03:09:06 & AO/L-wide    & 1700  & 0 \\
2016-01-04 & 03:40:06 & AO/L-wide    & 1200  & 0 \\
2016-01-07 & 23:14:07 & GBT/S-band   & 3300  & 0 \\
2016-01-08 & 00:35:28 & GBT/820\,MHz & 1513  & 0\\
2016-01-09 & 22:16:19 & GBT/S-band   & 3300  & 0 \\
2016-01-11 & 00:57:08 & GBT/S-band   & 3300  & 0 \\
2016-01-11 & 02:06:44 & GBT/820\,MHz & 2264  & 0\\
2016-01-15 & 00:54:35 & Lovell       & 7208 & 0 \\
2016-01-20 & 02:43:36 & Lovell       & 7215 & 0 \\ 
2016-01-23 & 00:56:47 & AO/L-wide    & 6485 & 0 \\
2016-01-26 & 00:58:17 & AO/L-wide    & 6058 & 0 \\
2016-01-31 & 00:28:44 & AO/L-wide    & 6362 & 0 \\
2016-02-01 & 00:30:20 & AO/L-wide    & 6284 & 0 \\
\hline
\end{longtable}
\end{table}

\subsection{Jansky Very Large Array}
\label{sec:vlaobs}

We observed the \frb\ field at 1.6\,GHz for 10\,hr with the Karl G. Jansky 
Very Large Array (VLA) in D-configuration (Project Code: VLA/15B-378) 
to better localize the FRB position
and to set limits on the distribution of free electrons along the line of
sight (see Section \ref{sec:hii}). The 10\,hr were split into ten 1-hr
observations occurring every few days from 2015~November~25 to 2015~December~17.  
Each 1-hr observation consisted of a preliminary scan of the flux 
calibrator J0542$+$4951 (3C147) followed by three 14-min scans on the 
FRB field bracketed by 100-s scans on the phase calibrator J0555$+$3948.

Data were collected in the shared-risk fast-sample correlator mode 
\citep[see, e.g.,][]{lbb15} with visibilities recorded every 5\,ms.  
The bandwidth available in this mode is currently limited by the correlator 
throughput to 256\,MHz, which we split into two 128-MHz subbands.  The 
subbands were centered on frequencies of 1435.5\,MHz and 1799.5\,MHz 
to avoid known RFI sources.  By observing in the fast-sample mode, 
we are able to produce channelized time series data for any synthesized beam within 
the $\sim 0.5^{\circ}$ primary field of view (28\arcmin\ across)
in addition to the standard interferometric visibilities. 
If a pulse is detected in the time-series data, it will localize \frb\ to
within one synthesized beam (about $30\arcsec-45\arcsec$ in D-configuration at 1.6\,GHz,
depending on hour angle coverage and visibility weighting in the image).
The results of the time-domain burst search will be presented in a subsequent paper
(Wharton et al., in prep).  The analysis of these data is described below in Section \ref{sec:vla}.  Here we present only the imaging results.

\subsection{\swift\ \& \cxo}
\label{sec:xrayobs}

We performed observations with the \swift\ X-ray Telescope \citep{bhn+05},
which is sensitive to X-rays between 0.3--10\,keV, 
on 2015 November 13, 18, and 23 
(Obs IDs 00034162001, 00034162002, 00034162003).
The observations were performed in Photon Counting (PC) mode, which
has a time resolution of 2.5\,s and
had exposure times of 5\,ks, 1\,ks and 4\,ks, respectively.
We downloaded the Level~1 data from the HEASARC archive and
ran the standard data reduction script {\tt xrtpipeline}
using HEASOFT~6.17 and the \swift\ 20150721 CALDB.

On 2015 November 23, \cxo\ X-ray Observatory observations were performed 
using ACIS-S in Full Frame mode, which provides a time resolution of
3\,s and sensitivity to X-rays between 0.1--10\,keV (Obs ID 18717). 
The total exposure time was 39.5\,ks. 
The data were processed with standard tools from
CIAO~4.7 and using the \cxo\ calibration database CALDB~4.6.7.
Note that we also performed a simultaneous 1.5-hr GBT observation during the 
\cxo\ session (see Section \ref{sec:gbtobs}) but detected no radio bursts
in those data.

For both the \swift\ and \cxo\ observations,
we corrected the event arrival times to the solar system
barycenter using the average \src\ position from \citet{ssh+16}.
The analysis of these X-ray observations is described below in Section 
\ref{sec:xrayanalysis}.

\section{Repeating Radio Bursts}
\label{sec:bursts}

\subsection{Burst Search}
\label{sec:burstsearch}

To search for bursts from \src, we used standard tools from the PRESTO software
suite \citep{ran01}\footnote{\url{http://www.cv.nrao.edu/~sransom/presto/}}. 
We first identified RFI-contaminated frequency channels and 
time blocks using {\tt rfifind}. Those channels and blocks were masked in subsequent
analysis. 
We performed two searches: we first searched using the full instrumental time
resolution (see Section \ref{sec:obs}) in a narrow DM range and then a
coarser search where the data were downsampled to 163.84\,$\micro$s in a
DM range of 0--10000\,\dmunits.

We performed a search for bursts in the DM range of 0--10000\,\dmunits\ on
data that were downsampled to 163.84\,$\micro$s.
Dedispersed time series were produced with a step size depending on the 
trial DM and selected it to be optimal given the amount of interchannel smearing 
expected based on the time and frequency resolution of the data.
For each time series, we searched for significant single-pulse signals 
using {\tt single\_pulse\_search.py}. 
For Effelsberg, Jodrell, and GBT observations we have searched down
to a signal-to-noise threshold of seven. For Arecibo, which suffers from
much stronger and persistent RFI, we have an approximate S/N$\sim$12 threshold.
More sophisticated RFI excision could lead to the identification 
of additional, weak bursts in these data sets.
A deeper search of the data can also be guided by the possible future 
determination of an underlying periodicity.

\begin{figure*}
\plotone{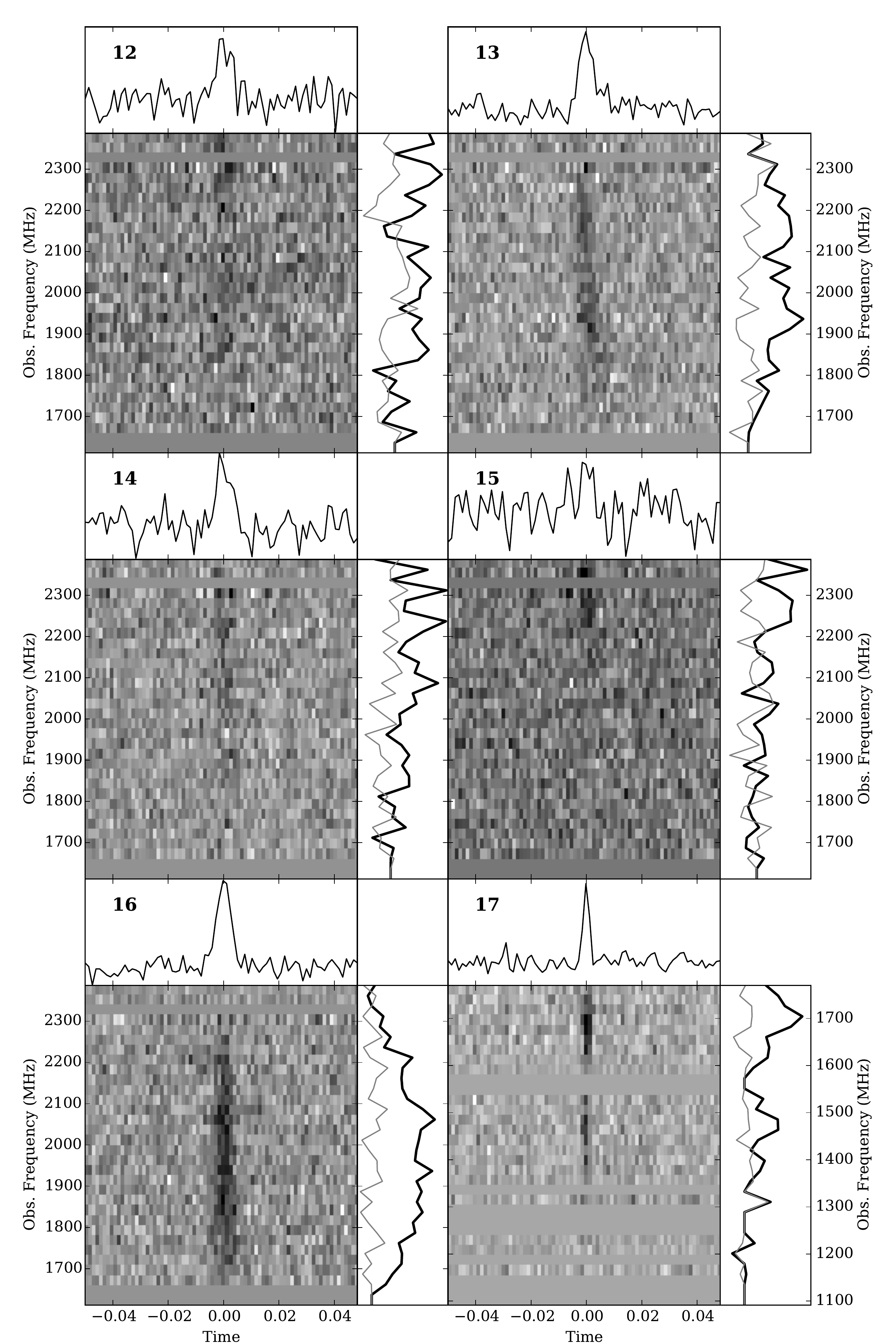}
\figcaption{
Dynamic spectra for each of the bursts detected in 2015 November and December
using GUPPI (bursts $12-16$) and PUPPI (burst 17) dedispersed at DM=559\,\dmunits. 
For each burst, total intensity is shown in grayscale,
the top panels show the burst time series summed over frequency, and the 
side panels show bandpass-corrected burst spectra summed over a 10-ms
window centered on the burst. The on-burst spectrum is shown as a black
line and an off-burst spectrum is shown as a gray line to show the noise level.
Note that some frequency channels are masked due to RFI.
\label{fig:bursts}
}
\end{figure*}

In addition to the 11 bursts detected using the Arecibo ALFA receiver and 
reported by \citet{sch+14,ssh+16}, we have detected a further six bursts: five 
in GBT GUPPI observations with the S-band receiver and one in an Arecibo PUPPI 
observation using the L-wide receiver (see Table \ref{tab:obs}). 
All five GBT bursts were found at 2\,GHz; no bursts were found in the GBT
820-MHz observations. Further, four of them were found within a $\sim 20$\,min period
on 2015 November 19. No bursts were found
at DMs significantly different from the DM of \src.

The detection of a burst using the L-wide receiver suggests that the 
source is localized within the beam width of the receiver. We therefore
quote two uncertainty regions: one conservative one with a diameter of 6\arcmin\
\citep[as in][]{ssh+16} and a 3.1\arcmin\ region corresponding to the L-wide
FWHM.

In Figure~\ref{fig:bursts} we show each burst as a function of observing
frequency and time. Each burst has been dedispersed to a DM of 559\,\dmunits. 
The data have been corrected for the receiver bandpass, which was estimated
from the average of the raw data samples of each channel. That average
bandpass was then median filtered with a width of 20 1.6-MHz channels to remove
the effects of narrow-band RFI.
Frequency channels identified as containing RFI by PRESTO's {\tt rfifind}
were masked. 
The data were downsampled to 32 frequency channels and a time resolution
of 1.3\,ms.
The top panel for each burst plot shows the frequency-summed time series and the 
side panel shows the spectrum summed over a 10-ms window centered on 
the burst peak. We continue the burst numbering convention used in \citet{ssh+16}
whereby bursts are numbered sequentially in order of
detection. The first GBT burst is thus designated ``burst 12'', 
as the bursts presented in \citet{ssh+16} were numbered from 1--11. 
We choose this approach in order to avoid conflict with the burst identifiers used in \citet{ssh+16}, but caution that
there may be weaker pulses in these data which can be identified by future, deeper analyses. 

In Figure \ref{fig:subs} we highlight frequency-dependent profile evolution in
bursts 8, 10 and 13. As observed in \citet{ssh+16}, the ALFA-detected
bursts 8 and 10 show evidence for double-peaked profiles. Here we show that
the double-peak behavior is apparent at high frequencies, but the two peaks
seem to blend into a single peak at lower frequencies. For burst 13,
which is detected in only the top three subbands shown in Figure \ref{fig:subs}, 
the burst in the 1.8--2\,GHz subband is wider than in the two higher frequency
subbands, causing a bias in the DM measurement (see below).
To our knowledge, this is the first time frequency-dependent profile evolution
(not related to scattering)
has been observed in an FRB.

\subsection{Temporal and Flux Properties}
\label{sec:fluxprop}

For each burst we measured the peak flux, fluence and burst width (FWHM).
Before measuring these properties we normalized each burst time series
using the radiometer equation where the noise level is given by
\begin{equation}
\frac{T_\mathrm{sys}}{G \sqrt{2 B t_\mathrm{int}}}\,, 
\end{equation}
where $T_\mathrm{sys}$ is the system temperature of the receiver, 20\,K for 
GBT 2-GHz observations and 30\,K for Arecibo L-wide;
$G$ is the gain of the telescope, 2\,K/Jy for GBT and 10\,K/Jy for Arecibo; 
$B$ is the observing bandwidth; and 
$t_\mathrm{int}$ is the width of a time-series bin.
The peak flux is the highest 1.3\,ms-wide bin in this normalized time series.
The fluence is the sum of this normalized time series.
To measure the width we fit the burst with a Gaussian model. 
The peak time is the best-fit mean in the 
Gaussian model.

In Table~\ref{tab:bursts} we show the above measured properties for each burst. 
The GUPPI bursts have peak flux densities in the range $0.02-0.09$\,Jy at 2\,GHz
assuming the bursts are detected on axis.
These are similar to the range of peak flux densities found for the
ALFA-detected bursts presented by \citet{ssh+16}, $0.02-0.3$\,Jy.
The PUPPI-detected burst had a peak flux density of 0.03\,Jy, again assuming a perfectly on-axis detection, 
similar to the faintest bursts seen by \citet{ssh+16}.

\subsection{Dispersion Properties}
\label{sec:dmprop}

We measured the DM of each burst except burst 15, which was detected over too 
narrow a frequency range to make a reliable DM measurement. 
For each burst, time series were generated in subbands by averaging over 
blocks of frequency channels. The number of subbands depended on the 
signal-to-noise ratio (S/N) of 
the burst, and subbands with too little signal (S/N $<2$) or too contaminated 
by RFI were excluded. Times of arrival (TOAs) for each frequency subband 
were calculated using a Gaussian template. The width of the Gaussian template 
was chosen to match the burst width measured from the whole band, as described above. 
TEMPO2\footnote{\url{http://www.atnf.csiro.au/research/pulsar/tempo2/}} 
\citep{hem06} was used to find the best-fit DM from the TOAs. 
The results are given in Table~\ref{tab:bursts}. 

\begin{figure*}
\plotone{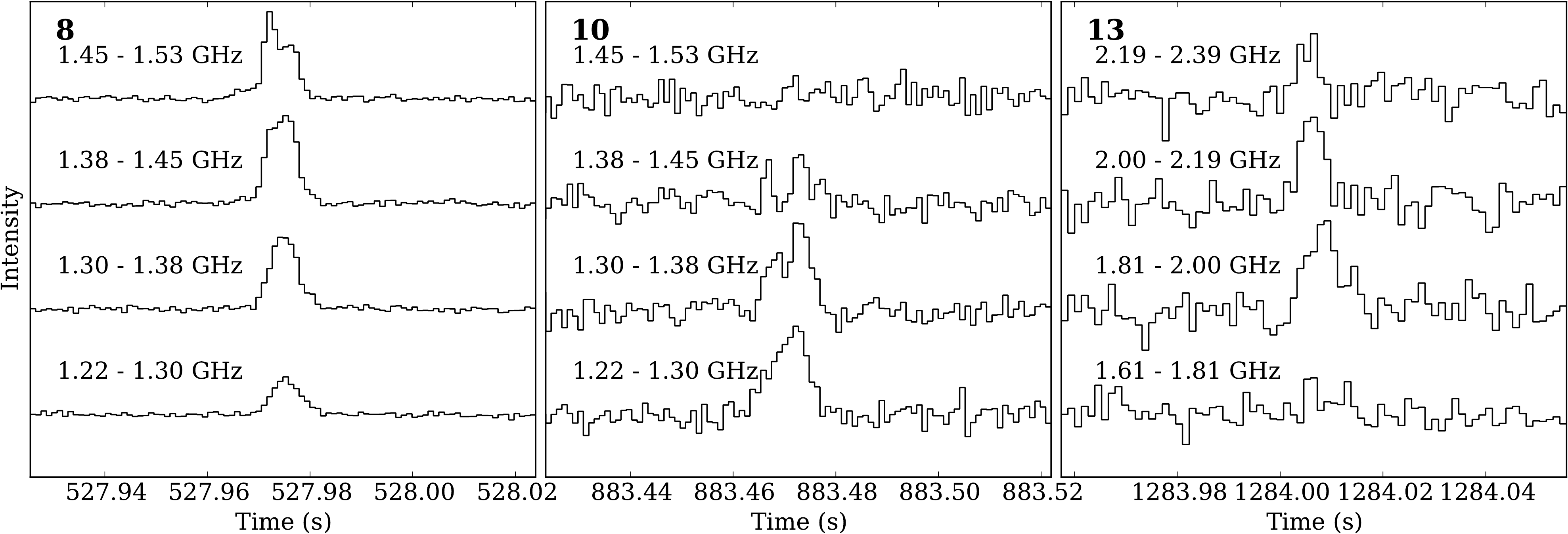}
\figcaption{
Burst time series plotted in four subbands for bursts 8, 10, and 13.
The frequency range used for each subband is indicated above each time series.
Frequency-dependent profile evolution is evident
(see Section \ref{sec:bursts}).
\label{fig:subs}
}
\end{figure*}

Several of these bursts (13 and 14) have formal DM measurements which are larger 
than the mean by a statistically significant margin. We argue that this 
reflects un-modeled frequency-dependent profile evolution (see Figure \ref{fig:subs})
and not a measurement of time-variable DM. 
Our simple Gaussian template assumes a constant 
burst shape and width across the band. 
If the burst shape varies across the band, then the TOA will shift to reflect the 
shift in the concentration of the flux density \citep[][]{hsh+12}. 

We explored this in depth for burst 13, which has higher S/N than burst 14. 
The value of 565.1$\pm$1.8\,\dmunits\ quoted in Table~\ref{tab:bursts} was 
calculated by dividing the full bandpass into eight subbands but only 
including the TOAs from the top five subbands in the fit 
(i.e.\ excluding data below 1900\,MHz due to low S/N). 
If instead we include TOAs above $\sim1700$\,MHz, the value drops to 560.0$\pm$1.2\,\dmunits. 
Similarly, calculating TOAs for four subbands instead of eight and including only the TOAs above 
1800 MHz gives a value of 569.2$\pm$1.8\,\dmunits. 
The spread in these fitted values is larger than the formal uncertainties 
reported by TEMPO2 by a factor of $\sim3$. 
Clearly there are systematic effects in the burst profiles which are not 
accounted for by the TEMPO2 uncertainties.

\citet{ssh+16} estimated the systematic uncertainty for bursts 1--11
by calculating the $\Delta$DM that results in a DM delay across the band equal 
to half the burst width. This has been calculated for each GUPPI and PUPPI-detected
burst and is the second uncertainty value given in Table~\ref{tab:bursts}.
Adding the systematic uncertainties to the statistical uncertainties brings 
the DMs into agreement at the level of 1.5$\sigma$.  Thus, there is currently no strong
evidence for variations in the DM between bursts.
The weighted average of the 15 bursts with measured DMs, excluding bursts 13 and 14
due to the above-mentioned profile variations, is $558\pm0.8$\dmunits.

Additionally, we measured the frequency index of the DM delay, 
i.e.\ $\Delta t_{\rm DM} \propto \nu^{-\xi}$, for the brightest of the GUPPI 
bursts (burst 16). 
We expect $\xi = 2$ for the propagation of radio waves through a cold, 
ionized plasma. 
The consistency of the observed frequency sweep of FRBs with $\nu^{-2}$ has been 
used to argue their astrophysical origins, as it would be highly unlikely that 
RFI would mimic the dependence so exactly\footnote{In fact, the `Perytons' 
\citep{bbe+11}, which have been shown to originate from on-site RFI at the 
Parkes observatory \citep{pkb+15}, do not 
precisely follow the expected pulse delay with frequency.}.
The DM index was measured using a least-squares 
fitting code, which is described in detail by \citet{ssh+16}. 
The measured index for burst 16 is $-1.997 \pm 0.015$, consistent with $-2$. 
Although the PUPPI-detected burst (burst 17) is also seen at high 
S/N, a large fraction of the band is masked due to RFI, making estimating
a dispersion index difficult.
The DM index has also been measured for 
both the discovery burst from \frb\ \citep{sch+14}, as well as the brightest of 
the bursts reported by \citet{ssh+16}, and all are consistent with $-2$. 

\subsection{Spectral Properties}

In Figure~\ref{fig:bursts}, we show the spectrum of each burst as a solid
black line in the right panel of each burst sub-figure. The gray lines
show the spectrum of the noise extracted from an off-pulse region that is
the same width as the on-pulse region (a 10-ms time window). Due to 
low S/N, it is difficult to say anything about the GUPPI-detected burst 
spectra except for the brightest bursts (bursts 13 and 16). 
Burst 17, the PUPPI-detected burst, is also seen at high significance, 
but much of the band is corrupted by RFI. 
Because of this, we do not attempt to fit any models to the spectra,
and only describe them qualitatively here. 
Both bursts 13 and 16 have bandwidths of $\sim600$\,MHz and drop 
off in S/N at the edge of the 1.6--2.4\,GHz band. 
It is clear that, as with the ALFA-detected bursts at 1.4\,GHz \citep{ssh+16}, 
the bursts detected at 2\,GHz are not well described by a broadband power-law 
spectrum and their spectra vary from burst to burst.

\begin{deluxetable*}{cccccc}
\tabletypesize{\small} 
\tablecolumns{8} 
\tablewidth{0pt} 
\tablecaption{ Burst Properties \label{tab:bursts}} 
\tablehead{ 
\colhead{No.}  & \colhead{Peak Time}  & \colhead{Peak Flux}     & \colhead{Fluence}  & \colhead{Gaussian}  & \colhead{DM}              \\
\colhead{}     & \colhead{(MJD)}      & \colhead{Density (Jy)}  & \colhead{(Jy\,ms)} & \colhead{FWHM (ms)} & \colhead{(pc\,cm$^{-3}$)} }
\startdata 
12 & 57339.356046005567 & 0.04 & 0.2 & $6.73\pm1.12$ & $559.9\pm3.4\pm3.7$\\              
13 & 57345.447691250090 & 0.06 & 0.4 & $6.10\pm0.57$ & $565.1\pm1.8\pm3.4$\\             
14 & 57345.452487925162 & 0.04 & 0.2 & $6.14\pm1.00$ & $568.8\pm3.2\pm3.4$\\              
15 & 57345.457595303807 & 0.02 & 0.08 & $4.30\pm1.40$ & -- \\            
16 & 57345.462413106565 & 0.09 & 0.6 & $5.97\pm0.35$ &  $560.0\pm3.1\pm3.3$\\ 
17 & 57364.204632665605 & 0.03 & 0.09 & $2.50\pm0.23$ & $558.6\pm0.3\pm1.4$ 
\enddata 
\tablecomments{Bursts 12--16 were detected at 2\,GHz at GBT and burst 17 was 
detected at 1.4\,GHz at Arecibo. The errors quoted on DM are, in order, 
statistical and systematic. Burst peak times are corrected to the solar
system barycenter and referenced to infinite frequency.} 
\end{deluxetable*}

\subsection{Polarization Properties}

Each GUPPI and PUPPI-detected burst was extracted
in the PSRFITS format using \texttt{dspsr}\footnote{\url{http://dspsr.sourceforge.net}}, 
retaining all
four Stokes parameters and the native time and frequency resolution of
the raw data.  These single pulses were calibrated with PSRCHIVE tools
using the pulsed noise diode and the quasar J1442+0958 as a
standard flux reference.  
No linear or circular polarization was
detected.  We searched for Faraday rotation using the 
PSRCHIVE\footnote{\url{http://psrchive.sourceforge.net}}
\texttt{rmfit} routine in the range ${\rm RM} \leq |20000|\,{\rm rad\,
  m^{-2}}$ but no significant RM was found.  It should be noted that
most of the GBT pulses are in a relatively low S/N regime, and a small
degree of intrinsic polarization cannot be ruled out.

\subsection{Periodicity Search}

In \citet{ssh+16}, we searched for an underlying periodicity in the burst 
arrival times by attempting to fit a greatest common denominator to the 
differences in time between the 8 bursts that arrived on 2015 June 2 
(bursts 4--11), but did not detect any statistically significant periodicities. 
Here, we apply an identical analysis to the four bursts detected on 
2015 November 19 (bursts 13--16) and found that it was not possible to 
find a precise periodicity that fit all of the bursts detected. 
We note that the minimum observed time between two bursts is 22.7\,s 
(between bursts 6 and 7), so if the source is periodic, the true period
must be shorter than this.
As we concluded from the 
ALFA-detected bursts, more detections are necessary to determine 
whether any persistent underlying periodicity to the bursts is present.

In addition, we conducted a search for a persistent periodicity 
on dedispersed Arecibo and GBT 
observations using a fast-folding algorithm 
(FFA)\footnote{Adapted from \url{https://github.com/petigura/FFA}}. 
Originally designed by \citet{sta69}, this algorithm operates in the 
time domain and is designed to be particularly effective at finding 
long-period pulsars \citep{lk05,kml+09}. 
The FFA offers greater period resolution compared to the FFT and has the 
advantage of coherently summing all harmonics of a given period, 
while the number of harmonics summed in typical FFT searches such as those 
performed by conventional pulsar search software, like PRESTO,
is restricted, typically to $\leq 16$. This makes the time-domain analysis 
more sensitive to low rotational-frequency signals with high harmonic content, 
i.e. those with narrow pulse widths,
which are often obscured by red noise \citep[e.g.][]{lbh+15}.
 
Prior to searching for periodic signals, RFI excision routines were applied to 
the observations. Such routines include a narrow-band mask generated by 
{\tt rfifind}, in which blocks flagged as containing RFI are replaced by 
constant data values matching the median bandpass. 
Bad time intervals were removed via PRESTO's clipping algorithm \citep{ran01} 
when samples in the DM = 0\,\dmunits\ time series significantly exceeded the 
surrounding data samples. Moreover, a zero-DM filtering technique as described 
in \citet{ekl09} was also applied to the time series by removing the 
DM = 0\,\dmunits\ signal from each frequency channel.
  
In this periodicity analysis, we searched periods ranging from 100\,ms 
to 30\,s.  Below 100\,ms, the number of required trials becomes prohibitive.
Re-binning was performed such that the FFA search was sensitive to all possible 
pulse widths, ranging from duty cycles of 0.5\% up to 50\%.
  
For every ALFA, PUPPI, and GUPPI observation 
(see Table \ref{tab:obs}), candidates were generated by the FFA 
from the dedispersed, topocentric time series. 
Approximately 50 of the best candidates for each time series were folded 
using PRESTO's {\tt prepfold} and then inspected by eye. 
No promising periodic astrophysical sources were detected.

\section{Multi-wavelength Follow-up}
\label{sec:multi}

\subsection{VLA Imaging Analysis}
\label{sec:vla}

The VLA data were processed with the Common Astronomy
Software Applications \citep[CASA;][]{mws07} package using the computing cluster
at the Domenici Science Operations Center in Socorro, New Mexico.
For the imaging analysis presented here, each observation was downsampled 
to increase the sampling time from 5\,ms to 1\,s. We then flagged the data and performed 
standard complex gain calibration of the visibilities using our flux and 
phase calibrators. Using the seven brightest sources in the field, we ran three 
rounds of phase-only self-calibration followed by one round of amplitude 
self-calibration.

The flagged and calibrated data were imaged using CLEAN deconvolution \citep{sch84}.
The two brightest sources in the field are located beyond the half-maximum point  
of the primary beam at 1.6\,GHz.  Since the primary beam width gets narrower with 
increasing frequency, these sources have very steep apparent spectral indices.
To account for the spectral shape of these sources, we used the multi-frequency
synthesis mode of the CASA CLEAN implementation \citep{rc11} and approximated 
the sky brightness as a first-order polynomial in frequency.  As a result, we get 
both an image and a spectral index map for each of the ten single-epoch observations.

We combined all the single-epoch observations into one multi-epoch data set and 
performed a single round of amplitude-only self-calibration to correct any amplitude 
offsets in the visibility data between scans.  The combined data were then imaged 
using the same procedure as the single-epoch imaging, producing by far the deepest radio continuum image 
to date for this field (see Figure \ref{fig:images}a).  
No obvious new sources (see below) are seen in our FRB detection overlap 
region. The central $5\arcmin \times 5\arcmin$ region of the image has an 
RMS noise of $\sigma = 60~\rm \micro Jy~beam^{-1}$, with pixel values ranging 
from $-156~\rm \micro Jy~beam^{-1}$ to $209~\rm \micro Jy~beam^{-1}$.  These results 
set a 5$\sigma$ upper limit on the flux density of any point sources of 
$S_{\rm max} = 0.3~\rm mJy$, a factor of $\sim10$ deeper than previous images
of the field \citep[from the NRAO VLA Sky Survey, NVSS,][]{ccg+98}.

While no obvious new sources fall within the FWHM of the two ALFA beams of the 
re-detections, there are two previously identified sources within the 
28~sq.\,arcmin conservative uncertainty region: 
J053210+3304 
($\alpha = 05^{\rm h}32^{\rm m} 10{.\!}^{\rm s}08(3)$, $\delta = 33^\circ 04\arcmin 05\farcs5(3)$)
and J053153+3310 
($\alpha = 05^{\rm h}31^{\rm m} 53{.\!}^{\rm s}91(2)$, $\delta = 33^\circ 10\arcmin 20\farcs2(2)$).
These sources were termed VLA1 and VLA2 by \citet{kon15}, who presented an 
analysis of archival data from NVSS \citep{ccg+98}.
J053210+3304 is consistent with a two-component AGN with the brighter component having a multi-epoch 
average flux density of $S_1 = 3.2\pm 0.1~\rm mJy$ and spectral index $\alpha_{1.6} = -1.1 \pm 0.1$.  
J053153+3310 has a multi-epoch average flux density of $S_2 = 3.0\pm 0.1~\rm mJy$ and spectral index of 
$\alpha_{1.6} = +1.7 \pm 0.1$.  
Imaging each of the ten epochs separately, we see that J053153+3310 is variable on timescales of 
a few days to a week, which is consistent with typical AGN variability 
timescales and amplitudes \citep[e.g.,][]{ofb+11}.
No other transient sources were detected in the single epoch images within the 
conservative positional uncertainty region for the FRB.

\subsection{X-ray Data Analysis}
\label{sec:xrayanalysis}

\begin{figure*}
\includegraphics[width=\textwidth]{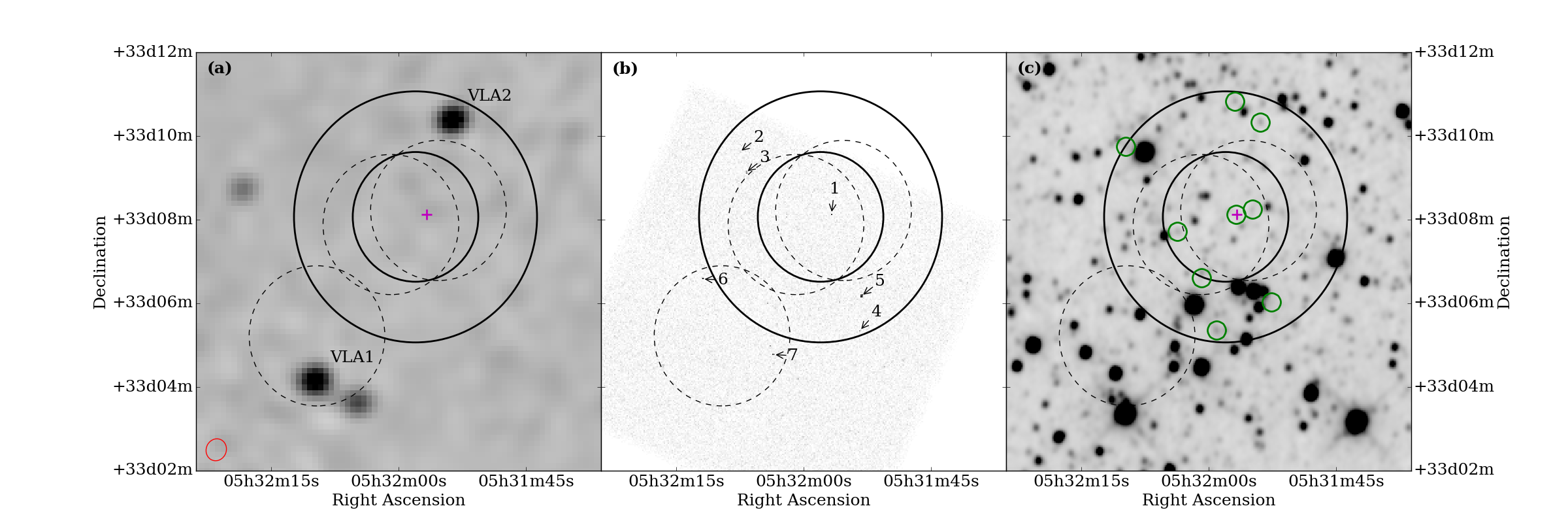}
\figcaption{
Radio, X-ray, and IR images of the FRB field.
In all panels: dashed circles show the ALFA beam locations in which bursts were 
detected \citep{sch+14,ssh+16}. 
Their diameters are the 3.4\arcmin\ FWHM of the ALFA beams.
The solid black circles denote the best estimated radio position with two uncertainty
regions shown: the 3.1\arcmin\ FWHM of the L-wide receiver (with which we have detected a burst) 
and the more conservative 6\arcmin\ diameter region.
(a) Jansky VLA 1.6-GHz image of the field of \src. 
The red ellipse (lower left) shows the approximate VLA synthesized beam size ($31\arcsec\times36$\arcsec).
Sources labelled ``VLA1'' and ``VLA2'' are the sources
J053210+3304 and J053153+3310 \citep[][see Section \ref{sec:vla}]{kon15}.
(b) \cxo\ image of the field of \src. The detected X-ray
sources are numbered as in Table \ref{tab:chandra}. 
(c) WISE 4.6-\micron\ image from the WISE archive. Green circles mark the 
position of the nine IR sources within the 6\arcmin\ diameter uncertainty region
that are identified as galaxies based on their WISE colors (see Section \ref{sec:OIR}).
In panels (a) and (c), the magenta cross represents the position of 
\chandrasrc\ which is numbered ``1'' in panel (b) and Table \ref{tab:chandra}.
\label{fig:images}
}
\end{figure*}

We searched the \cxo\ image (Figure~\ref{fig:images}b) 
for sources using the CIAO tool
{\tt celldetect}. Table \ref{tab:chandra} shows the coordinates
and number of counts in a circular extraction region with a 1\arcsec\ radius
($\sim90$\% encircled power)
for each detected source.
There are five sources within the conservative 6\arcmin\ diameter uncertainty region
shown as a solid circle in Figure~\ref{fig:images}b.
Only one of these X-ray sources, \chandrasrc\ (No. 1 in Table~\ref{tab:chandra} and               
Figure~\ref{fig:images}b) is within the
3.1\arcmin\ 1.4-GHz beam FWHM of our Arecibo PUPPI burst detection 
(see Section \ref{sec:burstsearch}).
We also searched the \swift\ observations using the HEASARC tool {\tt ximage} 
and found no sources within the more conservative 6\arcmin\ uncertainty region.

We cannot assume that \chandrasrc, or any of the other four sources
in the more conservative uncertainty region, is a counterpart of \src.
The \cxo\ image has seven detected sources within a 64\,sq.\,arcmin field of view.
Given this source density, the number of expected sources in any given
9\,sq.\,arcmin (the FWHM area of an Arecibo 1.4-GHz beam) region is $\sim1$.

If we assume that \chandrasrc\ is associated with \src, and that the
hydrogen column density, N$_\mathrm{H}$, is correlated with DM according to the 
prescription of \citet{hnk13}, then the N$_\mathrm{H}$ toward the source would 
be $1.7^{+0.7}_{-0.5}\times10^{22}$\,cm$^{-2}$ 
\citep[significantly higher than the predicted maximum Galactic N$_\mathrm{H}$ 
of $4.9\times10^{21}$\,cm$^{-2}$,][]{kbh+05}.
Assuming this N$_\mathrm{H}$, the best-fit power-law index for
a photoelectrically absorbed power-law model of the X-ray spectrum
is $3.3\pm0.4^{+0.7}_{-0.5}$ and the 1--10\,keV absorbed flux is 
$(9\pm3^{+0.7}_{-0.5})\times10^{-15}$\,erg\,s$^{-1}$\,cm$^{-2}$,
where the first errors on the index and flux are statistical uncertainties
and the second are systematic uncertainties from the spread in 
the N$_\mathrm{H}-$DM relation of \citet{hnk13}. 
We note that these values are heavily dependent on the assumed
N$_\mathrm{H}$.

We also searched for variability during the 39.5-ks observation
by making time series for each detected source with time resolutions
of 3, 30, and 300\,s. We then compared the predicted number of
counts in each time bin with the average count rate. No significant 
deviations from a Poisson count rate were found for any of the 
sources.

\begin{deluxetable*}{cccccccc} 
\tabletypesize{\normalsize} 
\tablecolumns{5} 
\tablewidth{0pt} 
\tablecaption{ Detected \cxo\ X-ray Sources \label{tab:chandra}} 
\tablehead{ 
\colhead{No.}  & \colhead{Name} & \colhead{RA (J2000)}      & \colhead{Dec (J2000)}    & \colhead{Counts}              & \colhead{Separation\tablenotemark{a}} & \colhead{IPHAS}       & \colhead{WISE} \\ 
\colhead{}     & \colhead{}     & \colhead{(hh:mm:ss.sss)}   & \colhead{(dd:mm:ss.ss)}  & \colhead{in 1\arcsec\ circle} & \colhead{(\arcmin)}      & \colhead{Counterpart} & \colhead{Counterpart} }

\startdata 
1 & CXOU~J053156.7+330807 & 05:31:56.711 & +33:08:07.59 & 35  & 0.28 & N & Y \\ 
2 & CXOU~J053207.5+330936 & 05:32:07.534 & +33:09:36.87 & 10  & 2.5  & Y & Y \\ 
3 & CXOU~J053206.8+330907 & 05:32:06.818 & +33:09:07.40 & 37  & 2.1  & N & N \\
4 & CXOU~J053153.4+330520 & 05:31:53.407 & +33:05:20.41 & 18  & 2.9  & N & N \\
5 & CXOU~J053153.2+330610 & 05:31:53.221 & +33:06:10.26 & 112 & 2.1  & N & N \\
6 & CXOU~J053211.9+330635 & 05:32:11.942 & +33:06:35.12 & 42  & 3.3  & N & N \\
7 & CXOU~J053203.5+330446 & 05:32:03.587 & +33:04:46.65 & 36  & 3.5  & Y & N 
\enddata 
\tablenotetext{a}{Angular offset from the average ALFA position of 
RA=$05^h31^m58^s$ and Dec=$+33\arcdeg08\arcmin04\arcsec$ (see Section \ref{sec:obs}).}
\end{deluxetable*}

\subsection{Archival IR-Optical Observations}
\label{sec:OIR}

The field of \src\ is in the anti-center region of the Galactic plane, 
and has been covered by several optical and infra-red surveys. 
We have examined archival images from the {\em Spitzer GLIMPSE 360} survey, 
the {\em 2-Micron All-Sky Survey} (2MASS), and the {\em Second Palomar Observatory 
Sky Survey} (POSS II), among others. 
Due to the large beams of single-dish radio telescopes in comparison to 
the density of sources found in
optical and infra-red surveys, there are many sources within the positional uncertainty
region of \src. 

The {\em Wide-field Infrared Survey Explorer} \citep[WISE;][]{wem+10} space 
telescope covered the entire sky at multiple infrared bands 
(3.4, 4.6, 12, and 22\,$\micro$m wavelengths). 
Several objects are present in the 3.4 (W1), 4.6 (W2) and 12\,$\micro$m
(W3) images of the field around \src. The ALLWISE catalog \citep{cut13}
lists 182 sources consistent with the best position of \src\ (6\arcmin\
diameter), of which 23 have detections in the W1, W2 and W3
bands. \citet{nhn+14} find that the W2$-$W3 color
is a reliable indicator to distinguish between stars and galaxies. They suggest
using the criterion W2$-$W3 $> 2$ for separating galaxies from stars. Using this
indicator, we find that 9 of the 23 objects can be classified as being
galaxies. 
In Figure \ref{fig:images}c we show the 4.6-\micron\ (W2) image and 
mark the 9 sources identified as galaxies.
We treat this number of galaxies as a lower limit, as many WISE sources
are not detected in the W3 band. Further, there are undoubtedly many galaxies
that would not be detected in any WISE band.
For this reason it is extremely difficult to identify a host galaxy without further 
localization of the FRB source.

Alternatively, in order to further investigate the possibility of a Galactic 
origin, here we report on the two most constraining archival observations 
in terms of testing for the existence of a hypothetical Galactic \ion{H}{2} 
region that could provide the excess dispersion of \frb\ compared with
the maximum line-of-sight value expected from the NE2001 model.

The WISE 22-$\micro$m band images, with 12\arcsec\ resolution and a sensitivity of 
6\,mJy, have been used by \citet{abb+14b} to catalog Galactic 
\ion{H}{2} regions with a high degree of completeness in conjunction with 
radio continuum imaging. We have extracted WISE 22-$\micro$m data in the 
region of interest from the NASA/IPAC Infrared Science Archive and find no 
sources within \frb's positional uncertainty 
region, down to the sensitivity limit of the \citet{abb+14b} survey.

The {\em Isaac Newton Telescope Photometric H-Alpha Survey} 
\citep[IPHAS;][]{dgi+05,bfd+14} provides optical survey images of the Galactic plane 
in our region of interest with a median seeing of $1.2^{\prime \prime}$ and a broadband 
magnitude limit of $\sim20$ or better at SDSS $r'$~band, as well as narrow-band 
H$\alpha$\ observations that are typically a magnitude brighter in limiting 
sensitivity. Since the field of interest is in the Galactic anti-center, 
the images are relatively uncrowded for the low Galactic latitude, 
and the H$\alpha$\ image is devoid of any extended 
emission. No evidence is seen for a planetary nebula or a supernova remnant 
within the region.

In combination with our VLA 1.6-GHz (20-cm) observations, the non-detections 
at 22\,$\micro$m and H$\alpha$ bands place a stringent limit on any Galactic 
\ion{H}{2} regions or other sources that might contribute to \frb's high DM, 
as discussed further in Section \ref{sec:hii}
\citep[see also][]{kon15}.

We can also look for optical and IR counterparts of the 
\cxo\ sources (Table \ref{tab:chandra} and Figure \ref{fig:images}b)
in the WISE and IPHAS source catalogs. 
For each \cxo\ source, we looked for sources in the IPHAS catalog
that are within 2\arcsec\ of the \cxo\ position and for sources within
10\arcsec\ from the WISE catalog. In Table \ref{tab:chandra} we show
whether each source has an IPHAS or WISE counterpart. The source 
CXOU~J053207.5+330936 has both an IPHAS and WISE counterpart and is
coincident with the star TYC~2407-607-1 \citep{hfm+00}.
The source CXOU~J053203.5+330446 has an IPHAS optical counterpart ($r'=15.6$, $i'=15.0$)
but no WISE counterpart (note that this source is outside our conservative
uncertainty region). The source \chandrasrc, the only source within the 1.4\,GHz
FWHM region of \src, does not have an optical 
counterpart in IPHAS, but does have a WISE counterpart (one of the 9 WISE sources
classified as galaxies above as shown in Figure \ref{fig:images}c).
Given this galaxy classification and the fact that the vast majority of faint 
\cxo\ X-ray sources are AGN \citep{bab+04a} this source is most likely an AGN.
Indeed, the expected number of AGN in the FWHM region of \src\ at the X-ray
flux level of \chandrasrc\ is $\sim$1 \citep{bab+04a}.
Note that at this point the suggested AGN nature of this source does not
preclude it from being the host galaxy of \src.

\subsection{Archival High-Energy Observations}

An examination of the transient catalogs from \textit{Swift} BAT, 
\textit{Fermi} GBM, {\em MAXI}, and {\em INTEGRAL} reveals that no hard X-ray/soft 
$\gamma$-ray bursts have been reported within $\sim1\arcdeg$ of \frb's position. 
In principle, high-energy counterparts to the radio bursts may be below the 
significance threshold necessary to trigger a burst alert. 
To explore this possibility, we retrieved the \textit{Fermi} GBM daily event 
data from all twelve detectors to check for any enhancement in count rate 
during the times of the FRB bursts. We find that the event rates within 
$\pm$10\,s of the radio bursts (after correcting for dispersive delay between 
radio and gamma-ray arrival times) are fully consistent with that of the 
persistent GBM background level.
The absence of soft $\gamma$-ray emission associated with the \src\ 
radio bursts rules out a bursting magnetar within a few hundred kpc \citep{ykh+16}.

In the \textit{Fermi} LAT 4-year Point Source Catalog \citep[3FGL;][]{aaa+15} 
there are no $\gamma$-ray sources positionally coincident with \frb. 
This is confirmed by a binned likelihood analysis, which shows no $\gamma$-ray 
excess at \frb's position that may arise due to a persistent source.  
This is not surprising, given the Galactic latitude of $b = -0.2^{\circ}$, 
where the diffuse $\gamma$-ray background is high. On the other hand, 
if the $\gamma$-ray emission is also transient, the source may be evident 
in the \textit{Fermi} LAT light curve. To this end, we generated 
exposure-corrected light curves using aperture photometry with a circle of 
radius 1$^{\circ}$ binned at 1, 5, 10, and 60-day intervals. 
None of these exhibit any statistically significant increase in flux, 
including around the times of the radio detections. 
In addition, within 1$^{\circ}$ of \frb, 
none of the individual $\gamma$-rays detected arrive within 10\,s of the 
arrival time of the radio bursts. 

\section{Galactic or Extragalactic?}
\label{sec:hii}

It is important to consider whether \frb\ may be Galactic, despite
its DM being three times the maximum Galactic contribution predicted along the line of sight.
The source's repetition sheds new light
on this question.  Here we make use of some of the arguments
of \citet{kon15}, who gave this source considerable thought prior to
our discovery of repeat bursts.

Assuming a Galactic contribution of 188\,\dmunits\ \citep{cl02} to the DM of
559\,\dmunits\ for \src, we take the `anomalous' amount that
must be explained to be DM$'\equiv 559-188=371$\,\dmunits.

This anomalous DM contribution could be explained by an intervening ionized 
nebula aligned by chance along the line-of-sight, or one in which the source 
is embedded.  As discussed in
\citet{kon+14} and \citet{kon15}, such a nebula will also necessarily
emit and absorb radiation.  We now show that, under reasonable
assumptions, such emission should have been detected if the source is located within our Galaxy.  
Here DM$'= n_e L_{pc}$ for a homogeneous electron distribution
and $L_{pc}$ is the putative nebular size in pc.
Such a nebula has an emission measure EM=DM$'^2$/$L_{pc}
= 138,000 L_{pc}^{-1}$~pc~cm$^{-6}$ if one ignores electron
density fluctuations in the nebula.
If fluctuations are included or if the filling factor $\phi$ is small the EM 
would be larger than assumed. However, such an increase would only make our limit
on a putative nebula more constraining.
In the following, we assume a spherical nebula
but address the validity of this assumption at the end.

Assuming a nebular electron temperature of 8000\,K, typical for 
such photoionization, the optical depth to 
free-free absorption is \citep[e.g.][]{kon+14}
\begin{equation}
\tau_\mathrm{ff} = 4.4 \times 10^{-7} {\rm EM} \left( \frac{T_e}{8000 \; {\rm K}} \right)^{-1.35} \left( \frac{\nu}{1 \; {\rm GHz}} \right)^{-2.1},
\end{equation}
or
\begin{equation}
\tau_\mathrm{ff} = 0.03 L_{pc}^{-1}  \left( \frac{T_e}{8000 \; {\rm K}} \right)^{-1.35} \left( \frac{\nu}{1.4 \; {\rm GHz}} \right)^{-2.1},
\label{eq:tauff}
\end{equation}
where the frequency normalization has been changed to be the approximate center of the ALFA band.

The highly variable spectra of the source in the ALFA band, as reported by \citet{ssh+16}, indicate
that they are intrinsic to the source and suffer little absorption, 
i.e. $\tau_\mathrm{ff} \ll 1$ (see Section \ref{sec:specshape}).  
Equation~\ref{eq:tauff} therefore implies
$L_{pc} \gg 0.03$\,pc.   
At a fiducial Galactic distance\footnote{For a fiducial distance to 
the source of 5\,kpc, the Galactic column would have to be significantly less 
than the maximum along this
line of sight, but this would only strengthen the conclusions that follow, since
the electron column of a putative nebula, DM$'$, would have to be even higher.} 
of 5\,kpc, such a source's angular extent $\theta \gg 1.2$\arcsec.
Given that DM$'= n_e L_{pc}$, 
this implies an electron density $n_e \ll 12,500$\,cm$^{-3}$ 
and emission measure EM=DM$'^2 / L_{pc} \ll 4.6 \times 10^6$\,pc\,cm$^{-6}$.

Note that the lack of deviation of the dispersion index from the cold plasma 
dispersion value of $-2$ also yields a constraint on source size \citep{katz14}.  
However it is far less constraining than that from the absence of free-free absorption.

In the optically thin regime, the free-free volume emissivity is given by 
$\epsilon_{\nu} = 5.4 \times 10^{-39} T_e^{1/2} n_e^2 g
$~erg~cm$^{-3}$~s$^{-1}$~Hz$^{-1}$~sr$^{-1}$ 
\citep{rl79} for a pure hydrogen
plasma, where $g$ is the Gaunt factor, approximately 5.5 for our case. 
Note that the above expression is roughly independent of $\nu$ in the optically thin regime.
Normalizing in temperature and using $n_e =$DM$'/L_{pc}$ we have
\begin{equation}
\epsilon_{\nu} = 4.5 \times 10^{-35} \left ( T_e / 8000 \; \rm{K} \right )^{1/2} L_{pc}^{-2} \;\; {\rm erg~cm}^{-3} \; {\rm s}^{-1} \; {\rm Hz}^{-1} \; {\rm sr}^{-1}.
\end{equation}
To get the luminosity density $l$ in erg~s$^{-1}$~Hz$^{-1}$, 
assuming isotropy, we multiply the above expression by $4 \pi$, 
as well as by the volume of the nebula, $V = (4/3) \pi (L_{cm}/2)^3$~cm$^3$,
and then divide $l$ by $4 \pi d_{cm}^2$ to get the total source flux density.
To compare with observed maps of the region, we must further multiply by
$(\theta / \phi_{beam})^2$ for an unresolved source, and by $(\phi_{beam}/ \theta)^2$
for a resolved source, where $\theta$ is the source angular extent on the 
sky given $L$, and $\phi_{beam}$ is the angular resolution of the map.

\begin{figure}
\plotone{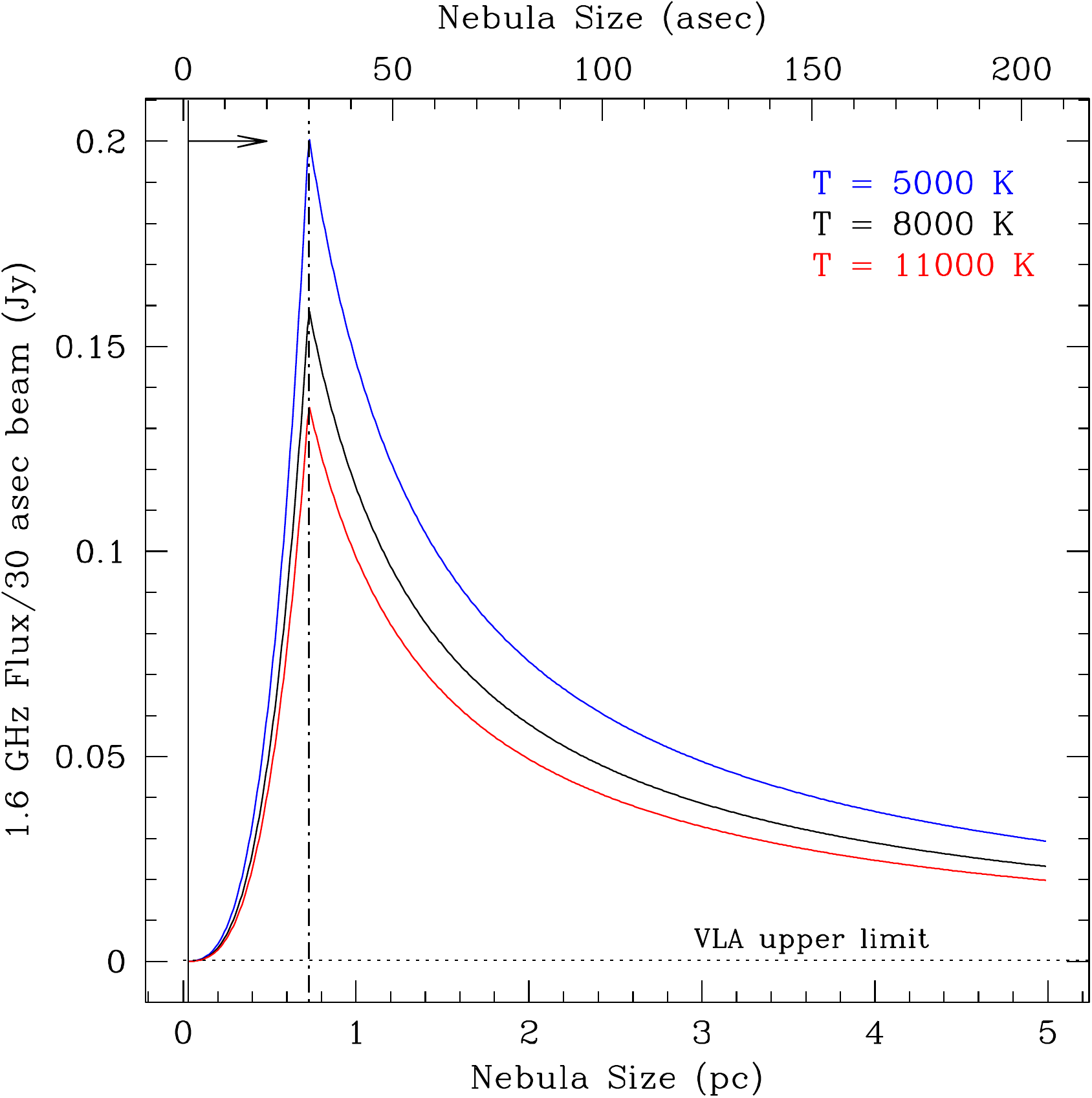}
\figcaption{
Predicted 1.6-GHz free-free continuum flux per 30$^{\prime \prime}$ beam (corresponding to the 
angular resolution of our VLA observations; Section \ref{sec:vla})
versus assumed size of a putative intervening
ionized nebula, in pc (lower horizontal axis), and in arcseconds (upper horizontal axis)
for a distance of 5\,kpc.  The vertical solid line shows the size constraint 
(lower limit) from the knowledge that the nebula must be optically thin.  
The vertical dot-dashed line
corresponds to the angular resolution of our VLA data.  The upper limit on point source
flux in our field is shown with a horizontal dotted line and corresponds to 0.3\,mJy/beam.
Curves for three different assumed plasma temperatures are shown.
\label{fig:vlaHII}
}
\end{figure}

As shown in Figure~\ref{fig:vlaHII}, from our VLA upper limit (0.3\,mJy/beam in a 
$\sim30\arcsec$ beam; see Section \ref{sec:vla}) 
we can rule out any such nebula based on the absence of free-free continuum emission at 20\,cm. 
This is consistent with the conclusion that there is no \ion{H}{2} region along the line of sight
from the WISE 22-$\micro$m map \citep[see Section \ref{sec:OIR} and][]{abb+14b}. 

Moreover, recombination radiation in the form of H$\alpha$ emission would also have
to be present for a putative nebula; next we show that the predicted H$\alpha$ flux 
is ruled out by observations in the IPHAS H$\alpha$ survey of the northern Galactic plane 
\citep{dgi+05}.
As discussed by \citet{kon+14} and \citet{kon15}, the H$\alpha$ surface brightness is given by
$I = 1.09 \times 10^{-7}$EM\,erg\,cm$^{-2}$\,s$^{-1}$\,sr$^{-1}$.  
If we take EM = DM$'^2$/$L_{pc}$ and integrate $I$ over the area of the nebula, $\pi \theta^2 /4$,
the permitted range of $L_{pc}$ implies a predicted H$\alpha$ flux of
$1.4 \times 10^{-11}$\,erg\,cm$^{-2}$~s$^{-1} < f_{H\alpha} < 1.2 \times 10^{-10}$\,erg\,cm$^{-2}$\,s$^{-1}$
or in IPHAS parlance \citep[see Eq. 11 and Appendix D in][]{kon15}
$8.0 < ha < 10.4$ for an assumed distance of 5\,kpc.  Even at a distance of 20\,kpc, $ha < 13.4$.
This is strongly ruled out given that the IPHAS catalog shows no non-stellar sources
in our field 
yet contains objects as faint as $ha = 19$ \citep{kon15}.

As argued by \citet{kon+14}, a nebula having electron density filling
factor $\phi < 1$ only strengthens the constraint on $L_{pc}$, making the lower limit
larger by a factor of $1/\phi$.  Also, those authors discussed the possibility of
a flash-ionized nebula but predict a flux level comparable to those estimated
above, and recombination timescale of years.  This is therefore ruled out for \src.  One possibility we cannot absolutely exclude, however, is a greatly
elongated nebula with long axis specifically in our direction.  However, this seems contrived and highly unlikely.
Note that the above conclusions apply even if the 
\citet{cl02} NE2001 model under-predicts the Galactic column along this line-of-sight
by as much as a factor of 2. 

Further, there is no evidence that the NE2001 model severely underpredicts the 
DM in this direction in the pulsar population. As we stated in \citet{ssh+16},
the highest-DM pulsar known in a 20-degree radius around \src\ is the 
millisecond pulsar PSR~J0557+1550 \citep{skl+15} which has a DM of 103\dmunits,
60\% of that predicted for the max Galactic column along that line-of-sight.
The DM of \src\ is thus clearly an outlier, regardless of the accuracy of the
NE2001 model.

\section{Updated PALFA FRB Detection Rate}
\label{sec:rate}

Several studies have noted a possible dependence of the FRB rate on Galactic latitude \citep[e.g.][]{pvj+14}. 
One possibility, for an underlying source population that is isotropic, is 
that foreground effects such as unmodeled Galactic scattering \citep{bb14}
result in relatively fewer detections near the Galactic plane.   Alternatively,
\citet{mj15} have suggested that the rate of FRBs far out of the plane may be enhanced relative to that in the plane
due to diffractive scintillation effects.
The PALFA survey is in principle well suited to test these claims by comparing 
its FRB detection rate with that of higher latitude surveys that
also operate at 1.4\,GHz.  Here we arbitrarily define a FRB as any few-ms radio 
burst that has DM at least twice the maximum along the line of sight as 
predicted by the \citet{cl02} NE2001 model, and for which the DM excess
cannot be explained by any intervening Galactic structure.  

To estimate the FRB rate implied by the PALFA survey to date, 
we make use of the results of the PALFA analysis pipeline described by 
\citet{lbh+15} and run on the Guillimin supercomputer operated by Compute Canada and McGill University.  As part of this pipeline
PRESTO's routine {\tt single\_pulse\_search.py} was run on all PALFA data obtained after 2009 March 17 using the Mock spectrometers
\citep[see][for details]{lbh+15}.  This pipeline's output was subject to the grouping
and rating scheme {\tt RRATtrap} described by \citet{kkl+15}. 
In the PALFA analysis pipeline, any beam containing a single pulse having 
S/N greater than 9.2 was inspected.   
This threshold S/N applies for all the searched pulse widths, which range from 1 to 150 times
the input time series bin size. Due to downsampling in the optimal dispersion plan,
the time resolution ranged from 65.5\,$\micro$s at DM$<212.8$ and 2.0\,ms at 
DM$>3266.4$ \citep[see][for details on the dedispersion plan]{lbh+15}.
To convert this S/N to a flux density, we use
the expression given by  \citet{cm03}:
\begin{equation}
    S_i=\frac{ (S/N)_b S_{sys}}{W_i} \sqrt{ \frac{W_b}{n_p B}},
    \label{eq:spflux}
\end{equation}
where $S_i$  is the intrinsic flux density of the pulse, 
$(S/N)_b$ is the measured S/N of
the broadened pulse, $S_{sys}$ is the system-equivalent flux density,
$W_i$ and $W_b$  are the intrinsic and
broadened pulse widths, respectively, $n_p$  is the number
of summed polarizations, and $B$ is the bandwidth.
Here we adopt $(S/N)_b = 9.2$,
$W_i = W_b = 1$\,ms,  $n_p = 2$, and $B = 322$~MHz.
We consider two cases for $S_{sys}$: $S_{sys} = 5$~Jy, the field-of-view-averaged
system flux in the FWHM of the ALFA beams, 
and $S_{sys} = 27$~Jy for the full field-of-view of the sidelobe 
region down to the gain equivalent to that of the Parkes 1.4-GHz beam on-axis.
These numbers translate to a sensitivity limit of 57 mJy in the FWHM case 
and 300 mJy in Parkes-equivalent case.

As of 2015 November 28, PALFA has discovered $N_\mathrm{FRB} = 1$ \citep{sch+14} while surveying exclusively in the
Galactic longitude ranges $30^{\circ} < l< 75^{\circ}$ (inner Galaxy) and 
$174^{\circ} < l < 205^{\circ}$ (outer Galaxy)
and Galactic latitude range  $-5^{\circ} < b < 5^{\circ}$.   

The FRB rate can be determined using the total sky area surveyed per observation, $\Theta$, 
and the total time spent observing, $T_{obs}$.  
Here we ignore all observations having Galactic longitude $<40^{\circ}$ as 
these include lines of sight for which the 1.4-GHz scattering time is likely 
to be more than several milliseconds for sources ar large distances (\gapp\ 3\,kpc).
Then the rate, $R$, is given by 
\begin{equation}
    R = \frac{N_\mathrm{FRB}}{\Theta \times T_{obs}},
\label{eq:rate}
\end{equation}

Taking into account beam-dependent gain variations of
the ALFA receiver, each pointing covers $\Theta$=0.022\,sq.\,deg.~when 
considering the FWHM case (i.e. the region having at least half the
peak gain of the outer ALFA beams) 
and $\Theta$=0.105\,sq.\,deg.~when considering the 
Parkes-equivalent case \citep{sch+14}.  To determine
$T_{obs}$, we multiply the total number of pointings observed in the PALFA
survey between 2009 March 17 and 2015 November 28 by the average integration 
time per pointing, 233\,s. 
Also, we note
that any beam having more than 20\% mask fraction due to RFI 
went unanalyzed by our pipeline; these pointings
amount to 0.6\% of all beams and are ignored in this calculation.
Finally, we make a small correction for time lost due to RFI
masking. The average mask fraction for those observations 
that were analyzed by the pipeline (i.e. those with mask fraction $<20$\%) 
was 6\%, conservatively estimating that the 
entire fraction is from masking in time (rather than radio frequency).  
In this way, we find $T_{obs} = 36.9$\,days.
Using Equation~\ref{eq:rate}, we therefore find 
\begin{equation}
\begin{aligned}
R = 1.4\times 10^{-5}~\mathrm{FRBs~sq.~deg}^{-1}~\mathrm{s}^{-1}, \\ 
\mathrm{or}~5.1^{+17.8}_{-4.8} \times 10^4~\mathrm{FRBs~sky}^{-1}~\mathrm{day}^{-1}
\end{aligned}
\end{equation}
for the FWHM case, i.e. above 57\,mJy, and 
\begin{equation}
\begin{aligned}
R = 3.0 \times 10^{-6}~\mathrm{FRBs~sq.~deg}^{-1}~\mathrm{s}^{-1},\\ 
\mathrm{or}~1.1^{+3.7}_{-1.0} \times 10^{4}~\mathrm{FRBs~sky}^{-1}~\mathrm{day}^{-1}
\end{aligned}
\end{equation}
for the 
Parkes-equivalent case, i.e. above 300\,mJy,  
where we have assumed Poisson statistics to
evaluate the quoted 95\% confidence range.
Note that if we consider only the outer Galaxy
beams that PALFA has observed (where scattering is certainly minimal), 
our upper limits increase by approximately 65\% owing to reduced observing time.

Our estimated rate for the combined inner ($l > 40^{\circ}$) and outer Galaxy
survey regions is lower than the event rate estimated by
\citet{sch+14} because PALFA has observed for longer since the
latter calculation was done and \citet{sch+14} considered only
the outer Galaxy region.  Our number is 
consistent with the rate derived by \citet{rlb+16}, 
$4.4^{+5.2}_{-3.1} \times 10^3$~FRBs~sky$^{-1}$~day$^{-1}$, 
valid for a threshold
that is comparable to theirs. Note that although we have chosen to report a 
threshold based on minimum detectable flux for the Parkes gain averaged over
the beam, whereas \citet{rlb+16} report a fluence-complete limit for the 
boresight gain, these are minor differences in definition and our rates are
directly comparable.
However, do note that the Arecibo beam probes a deeper volume than the Parkes
beam due to its higher peak sensitivity. The rates thus may differ depending
on the underlying luminosity distribution of FRBs. Such a comparison is beyond
the scope of this paper.
Our low-latitude rate neither
confirms nor refutes the claimed dearth in the FRB rate at low Galactic
latitudes \citep{bb14}.

On the other hand, the above estimated PALFA FRB rate is in poor agreement with the prediction
of \citet{mj15} who assert that PALFA should detect $\sim 1$ FRB every four days
at low Galactic latitudes, based on a model in which diffractive interstellar
scintillation (DISS) is responsible for the disparity in event rates of FRBs between
high and low Galactic latitudes \citep{pvj+14}.  In the model, an effect associated
with Eddington bias enhances the rate of high-latitude FRBs since the expected
decorrelation bandwidth for DISS at high latitudes is comparable to the bandwidth.
But, at low latitudes, the narrower DISS bandwidth allows only small variations of the flux densities. 
In order to explain the disparity, \citet{mj15} require
a steep differential flux density distribution for the FRB
population ($p(S_{\nu}) \propto S_{\nu}^{-3.4}$ or steeper). This
implies a relatively large number of FRBs at low flux
densities, hence a higher predicted PALFA event rate, given PALFA's 
unparalleled raw sensitivity compared to other pulsar surveys. 
That we have detected only one FRB in $\sim 37$ days suggests 
that the difference in FRB rate at high and low latitudes is due primarily to an effect
other than that suggested by \citet{mj15}. Given the 95\% bounds on our updated rate
down to the Arecibo sensitivity, we infer a power-law index for the 
differential flux density distribution of $\alpha \gapp -2$.

\section{Discussion}
\label{sec:disc}

We have presented the discovery of six additional \frb\ bursts with GBT and Arecibo
as well as multi-wavelength images of the surrounding field.

The detection of \src\ with both Arecibo and GBT rules out a local source of 
RFI as the origin of the bursts (the telescopes are geographically separated 
by $\sim 2500$\,km).  As previously noted, the detection in only a single 
pixel of the Arecibo 7-beam ALFA receiver also shows that the bursts must 
originate beyond Arecibo's Fresnel length of $\sim 100$\,km \citep{kon15,ssh+16}.  
In a similar vein, it is also important to note that the consistent sky position 
of the bursts, to within at least $\sim 6^{\prime}$, indicates that the source 
at least approximately follows the sidereal reference frame and has a minimum 
distance of $\sim 1150$\,AU.  Hence, this also rules out the possibility that 
the bursts could originate from a man-made satellite. 

The bursts display extreme spectral variations and an episodic burst rate.
We have reiterated, using our new VLA images, as well as archival WISE and IPHAS data, 
that the source is almost
certainly extragalactic, as the high DM cannot be explained by any detected 
excess dispersive material within our Galaxy. 
Here we discuss the possible physical causes of the unusual properties of these
bursts in the context of an extragalactic origin and compare them to the other 
currently known FRBs.

\subsection{Spectral shape}
\label{sec:specshape}

The 11 bursts discovered in ALFA data presented by \citet{ssh+16} displayed
significant spectral variability. The bursts were observed with both
rising and falling spectra, and some were found with spectra peaking in 
the band. Overall, the spectral behavior could be described as consisting of a
feature with bandwidth comparable to the 322\,MHz ALFA band that varied
in peak frequency from burst to burst. 
It is not clear whether the behavior is 
caused by a single feature or a broadband spectrum that is 
strongly modulated at a frequency scale of 100--1000\,MHz.
The spectral behavior of the GBT bursts presented here in our 2-GHz observations
seems to be consistent with what we see at 1.5\,GHz, implying that the unusual spectral behavior
persists to at least the higher GBT frequency.

We can rule out DISS as the 
cause of the observed spectral structure on frequency scales 
$\Delta\nu \sim 600$\,MHz at both 1.5 and 2\,GHz.  
The frequency scale expected for DISS differs markedly from what is 
observed. The NE2001 estimate for the DISS bandwidth  
is only 57\,kHz at 1.5\,GHz and 200\,kHz at 2\,GHz, more than $10^3$ times 
smaller than the observed spectral structure (assuming an extragalactic origin).  
Unlike for DISS, the scale of the observed structure does not appear at least qualitatively 
different between 1.5\,GHz and 2\,GHz.   
In addition, the DISS time scale estimated by the NE2001 model \citep{cl02} is $\sim 4$\,min at
1.5\,GHz\footnote{The four minute estimate is based on the 
NE2001 model using an effective velocity of 100\,km\,s$^{-1}$ for changes 
in the line of sight to the source.  
However, for an extragalactic source the effective velocity is that of the
ISM across the line of sight, which is 
substantially smaller, leading to a longer DISS time scale.}, 
whereas we see extreme variation in spectral shape between
bursts separated by as little as 1\,min.

\newcommand{\thcrit}{\theta_{\rm crit}}

Empirically, we find no evidence for DISS in the burst spectra on {\it any}
frequency scale, provided the source is not nearby. 
Two effects can generally account 
for the absence of DISS: either finite source-size smearing of the diffraction
pattern which occurs at an angular size of $\gapp 1~\micro$\,arcsec at 1.5\,GHz, 
or the DISS bandwidth is too small to be 
resolved with our channel bandwidth $\Delta\nu_{\rm ch} = 0.33$\,MHz.   
However, for the source size to be $>1~\micro$\,arcsec, $\dnud$ would need 
to exceed $\Delta\nu_{\rm ch}$, which is only the case if the source is                                 
closer than about 2\,kpc (the distance at which the integration of the NE2001 model    
yields $\dnud = \Delta\nu_{\rm ch}$).    
Since the source appears to be unavoidably extragalactic,
as discussed in Section \ref{sec:hii}, it must be the case that 
$\dnud \ll \Delta\nu_{\rm ch}$.

It is also possible that the observed frequency structure is intrinsic to the source. 
Observations of giant pulses of the Crab pulsar at 1.4\,GHz have shown extreme 
spectral variability with spectral indices ranging from $-$15 to +15 \citep{kss10}.
At higher frequencies (5--10\,GHz), banded frequency structure has been observed
with similar bandwidths to those we observe in \src\ \citep[$\sim300-600$\,MHz;][]{he07}. 
The center frequencies of these Crab giant pulse bands also seem to vary
both during the microsecond duration pulses and from pulse to pulse.

\subsection{Episodic Behavior}

We have now observed \src\ for over 70\,hr using radio telescopes, 
with the majority of observations
resulting in non-detections (note, however, that the 70\,hr is spread over
telescopes with different sensitivities and these observations were performed 
at several different radio frequencies; see Table \ref{tab:scopes}). 
Of the 17 bursts found, six were found within a 
10-min period on 2015 June 2 and an additional four were found in a 20-min 
period on 2015 November 19. The arrival time distribution of the detected
bursts is therefore clearly highly non-Poissonian. We discuss here two 
possibilities for such variation: propagation effects due to the 
interstellar medium and variations intrinsic to the source.

DISS is highly unlikely to play any role in the intensity variations 
of the bursts from \src.
Bandwidth averaging of DISS yields a modulation index (RMS relative to mean intensity) that is only 
$m_d \sim 1 /\sqrt{0.3 B / \dnud} \sim 2.5$\% at 1.5\,GHz with the ALFA system (bandwidth 322\,MHz). 
The burst amplitudes therefore will be largely unaffected by DISS. 

However, refractive interstellar scintillation (RISS) 
may play an important role in the observed burst intensity 
variations on weeks to months time scales
as its characteristic bandwidth is $\Delta\nu_r \sim \nu \gg B$.   
Using expressions in \citet[][]{ric90},
we estimate the RISS modulation index to be  $m_r \sim 0.13$ based on the 
scaling for a Kolmogorov wavenumber spectrum for the electron density.  
We note, however, that measured modulation indices are often larger than the 
Kolmogorov prediction \citep{rl90,ks92}. 

The time scale for RISS is uncertain in this case, and the observed
episodic time scales are certainly within the range of these uncertainties. 
We estimate the characteristic time scale for RISS as follows.  
The length scales in the spatial intensity patterns of DISS and 
RISS, $l_d$ and $l_r$, are related to the Fresnel 
scale $r_{\rm F} = \sqrt{\lambda L / 2\pi}$ according to $l_r l_d = r_{\rm F}^2$,  
where $L$ is the effective distance to the Galactic scattering screen.  
In the NE2001 model, scattering in the direction of \src\ is dominated by the 
Perseus spiral arm at $L \sim 2$\,kpc,
giving $r_{\rm F} \sim 10^{11}$\,cm.  
The diffraction length scale is related to the DISS bandwidth as 
$l_d \sim \lambda \sqrt{d\dnud / 4\pi c}$ \citep[Eq.~9 of][]{cr98} from which 
we estimate  $l_r \sim 4 \times 10^{13}$\,cm. The corresponding time scale for 
RISS depends on the characteristic velocity for motion of the line of sight 
across the Galactic plasma. Using a nominal velocity of 100\,km\,s$^{-1}$, we obtain 
$\Delta t_r \sim 40$\,days.   
However, it is not clear what velocity to use because, unlike pulsars,  
the source motion is not expected to contribute.   If the motions are only from Galactic 
rotation with a flat rotation curve, the scattering medium and the Sun move 
together with zero relative velocity. However, non-circular motions, 
including the solar system's velocity relative to the local standard of rest
($\sim 20$\,km\,s$^{-1}$) and the Earth's orbital velocity, will contribute a 
total of a few tens of km\,s$^{-1}$ yielding an RISS time scale longer 
than 40 days.  Given the uncertainties in estimating RISS time scales, 
it is possible that $\Delta t_r$ ranges from tens of days to 100 days or 
more at 1.5\,GHz. RISS times scale with frequency as 
$\Delta t_r \propto \nu^{-2.2}$, so higher frequencies vary more rapidly.       

Although variability intrinsic to the source appears to dominate on minutes to hours
timescales, RISS could be important for intensity variations on timescales of
weeks to months. 
If we use $m_r = 0.13$ (allowing for the modulation index to be larger
than predicted by a factor of 2) at 1.5\,GHz and consider $\pm 2m_r$ variations, 
the maximum and minimum range of the RISS modulation is 
$g_{\rm r, max} / g_{\rm r, min} = 1.7$. If the burst amplitude distribution 
is a power law $\propto S^{-\alpha}$, as seen in the Crab pulsar's giant pulses 
\citep[for which $2.3 \lesssim \alpha \lesssim 3.5$; e.g.][]{mml+12},
RISS will cause the apparent burst rate above a detection threshold to vary by 
an amount  $(g_{\rm r, max} / g_{\rm r, min})^{\alpha-1}\sim 2-4$ for 
$\alpha = 2.3$ to 3.5. But note again that modulation indices have been 
observed to be higher than predicted, so the burst-rate variation could 
be still higher (e.g. for 
$m_r = 0.4, (g_{\rm r, max} / g_{\rm r, min})^{\alpha-1}\sim 20-200$).
Keeping in mind these uncertainties, it is clear that RISS
could cause large variations in the observed burst rate. This issue 
is being considered in detail in a separate article
(Cordes et al., in prep).

For intrinsic phenomena that might cause the episodic behavior, we look
at examples from pulsars within our Galaxy,
as supergiant pulses from pulsars and magnetars \citep{cw16,pc15} are
a viable model for repeating bursts.
First, the time separations in Crab giant pulses do not show a deviation
from a random process \citep{lcu+95}. However, a subset of young
Galactic pulsars display a phenomenon known as nulling in which they emit
pulses only a fraction of the time. The nulling fractions of these pulsars 
can be quite high, the most extreme emitting pulses $<$5\% of the time 
\citep{wmj07}.
Some Rotating Radio Transients (RRATs), a class of Galactic radio
pulsar that emit infrequent, bright millisecond duration 
radio pulses, also show episodic behavior in their pulses \citep{kle+10}. 
It is therefore possible that, if the bursts from \src\ originate
from a rotating neutron star, a similar process is causing the episodic
behavior in \src. 

Another example of young neutron stars in our Galaxy emitting
bursts in episodes are X-ray bursts from magnetars. Magnetar 
X-ray bursts have time scales from milliseconds to seconds and
are emitted in clusters during periods of outburst 
\citep[e.g.][]{gkw+01,gkw04,sk11}. \citet{lyu02} suggests 
that these X-ray bursts may have accompanying radio emission. To date, no radio
counterpart to a magnetar X-ray burst has been observed \citep[e.g.][]{tkp16}, but the 
possibility remains that it is what we are observing for \src.

\subsection{Comparison to other FRBs}

Here we compare some of the properties of \frb's 12 Arecibo-detected bursts 
against equivalent properties 
from the 15 FRBs so far reported from Parkes observations (see \pbj).
We limit ourselves to the 1.4-\,GHz detected bursts, so that
we are comparing bursts from the same frequency range to each other.

Based on the telescope gain and system temperatures of the Arecibo 
ALFA receiver and the Parkes multibeam receiver, Arecibo is about
a factor of 10 more sensitive to FRBs than Parkes (see Table \ref{tab:scopes})
at 1.4\,GHz. The limiting peak flux density (using a 1.3\,ms time scale as used 
in Section \ref{sec:fluxprop} to measure the peak fluxes of our bursts 
and a S/N ratio of five) of the Parkes multibeam 
receiver is 0.15\,Jy. Of our 12 1.4\,GHz Arecibo detections of \src, only burst 11
is brighter than this limit \citep{ssh+16}. So, if Parkes had undertaken a comparable 
follow-up campaign for FRB 121102, it may not have found repeat bursts.

The peak flux densities of all but one of the Parkes detected FRBs are
within an order-of-magnitude of the brightest \src\ burst, making the above 
comparison valid.  The notable
exception is FRB 010724 which is, with a peak flux density of $>30$\,Jy,
by far the brightest FRB detected to date
and, as it was the first discovered, the most followed-up with nearly a hundred
hours of telescope time spent checking for repeat bursts \citep{lbm+07}.
This seems to argue against the nature of FRB~010724 being similar to \src. 
Note however that the episodic behavior displayed by \src\ makes this 
comparison difficult. The underlying origin and timescales of this behavior
remains uncertain and the duty cycle could vary from source to source. 
If active periods of repeating FRB emission come and go, 
the follow-up observations of FRB~010724, performed six years after the initial burst,
could plausibly have resulted in non-detections if its periods of activity are
sufficiently rare.

We can also compare the widths of \src\ bursts to those of the Parkes FRBs.
Three of the Parkes FRBs, including the first reported \citep{lbm+07}, 
were detected with an analog filterbank system, with
3-MHz-wide frequency channels. These cause an intra-channel
dispersion smearing time, $\Delta t_{\rm DM}$, 8 times larger than
the equivalent for the other Parkes FRBs, detected with the BPSR backend
(with 0.39\,MHz channel width), or for the ALFA observations of
\src\ (0.34\,MHz width).

The 12 Arecibo \src\ pulses have measured widths spanning 3--9\,ms
\citep{ssh+16}. Given that $\Delta t_{\rm DM}$ for the bursts
detected with ALFA at 1.4\,GHz is 0.7\,ms and zero for the PUPPI-detected burst 
(as it was coherently dispersed), and that no scattering tail is
observed in the pulses, the observed widths must be the intrinsic 
width of the emission.

Two of the 15 Parkes FRBs have two-component profiles (as do some
of the \src\ pulses, although their shape is varied and not
obviously comparable to those of the Parkes events), and we neglect
them here.  The observed widths of the remaining 13 FRBs span
0.6--9\,ms, apparently comparable to the widths of \frb\ bursts. 
However, at least two \citep[see][]{tsb+13,rsj15},
and possibly four \citep[see][]{lbm+07,pbb+15}
of the Parkes FRBs display significant evidence of multi-path
propagation/scattering, which can make their observed widths
larger than their intrinsic widths. In addition, two Parkes FRBs have
$\Delta t_{\rm DM} = 7$\,ms, comparable to their observed widths
\citep{kkl+11,bb14}.

After accounting for all instrumental and measurable propagation
effects, none of the 13 Parkes single-component FRBs is 
wider than $\sim 3$\,ms. Indeed, several of the Parkes FRBs are 
temporally unresolved: e.g.,
FRB~130628 has an observed width of 0.6\,ms and $\Delta t_{\rm DM} \sim 0.6$\,ms; 
\citep{cpk+15}.

In summary, it appears that the intrinsic widths of the 12 \frb\ bursts
from Arecibo (3--9\,ms) are significantly longer than the intrinsic
widths of the 13 single-component Parkes FRBs ($\la 3$\,ms).

We can also compare the spectra of the \src\ bursts with those of
Parkes FRBs.
The collection of \src\ bursts has spectra that in some cases
can be approximated by power laws with indices spanning $-10 <
\alpha < +14$.  In some instances, the spectral behavior is more
complex and cannot be approximated by a power law \citep[][Fig. 2]{ssh+16}.

Only nine of the 15 Parkes FRBs have sufficiently well described spectral
properties to allow at least qualitative judgements on their spectra. In
most such cases, the original references do not provide quantitative
spectral information, but a ``waterfall'' plot (showing a grayscale
of the pulse flux density as a function of observing frequency, as in Fig. 2) is
available with sufficient S/N in these nine cases.
Note that the waterfall plots for FRBs~090625, 110703, and 130729 are not
published in the refereed literature but are available at FRBCAT.

For seven of the Parkes FRBs, within the available S/N, 
the spectrum appears consistent with showing
roughly monotonic flux density frequency evolution
and qualitatively consistent with mildly negative spectral indices
as for ordinary radio pulsars \citep[e.g.][]{mt77}. 
A closer look shows possible departures from
this general trend 
\citep[e.g., for FRB~110220; see Figure~S4 of][]{tsb+13}, 
but likely propagation effects need to be accounted
for when considering this issue in detail. Two Parkes FRBs have
published spectral indices: \citet{rsj15} report a marginally
inverted spectrum for FRB~131104, with $\alpha = 0.3\pm0.9$, but
caution that this value could be different depending on the true
position of the FRB within the telescope beam pattern.
\citet{kjb+16} report $\alpha = 1.3\pm0.5$ for FRB~150418, but
a similar caveat applies in that they assume the position of the
possibly coincident variable source detected in a galaxy within the
Parkes beam pattern \citep[but see][]{wb16,vrm+16}.

In summary, the largely qualitative spectra inferred for nine
Parkes FRBs may show in a few cases departure from standard pulsar-like
spectra, but even in the best such counter-examples, a possibly
uncertain FRB position renders such conclusions tentative.  In
contrast, the collection of Arecibo spectra from \src\
bursts displays intrinsic variability that includes examples with
very positive and very negative spectral indices, not represented
in the Parkes collection.

Two other useful quantities to compare are the observed peak flux
density $S_{\rm peak}$ and fluence $F$ of the various FRBs. Arecibo
detections of \src\ have $0.02 < S_{\rm peak} < 0.3$\,Jy and
$0.1 < F < 1$\,Jy\,ms. For the most part, the Parkes FRBs have
$S_{\rm peak}$ and $F$ an order of magnitude larger than the Arecibo
\src\ values: 0.2--2.2\,Jy and 1--7\,Jy\,ms, respectively. The
one exception is FRB~010724 \citep{lbm+07}, which is yet
another order of magnitude brighter ($S_{\rm peak} \sim 30$\,Jy and
$F \sim 150$\,Jy\,ms). 

The Arecibo ALFA system, used to detect the \src\ bursts near
1.4\,GHz, is $\sim 10$ times more sensitive than the multibeam
receiver system used to detect all Parkes FRBs. It is thus not surprising that the faintest Arecibo FRB
detections have flux densities an order of magnitude smaller than
those of the faintest Parkes FRBs. That the strongest
\src\ detections are one to two orders of magnitude below the
strongest Parkes FRB detections could reflect 
a luminosity distribution favoring fainter bursts.
Indeed, there are more \src\ bursts near the Arecibo detection
limit than high S/N bursts.
The lack of repeated burst detections to date from Parkes
FRBs might be due to the Parkes telescope's lower sensitivity not probing as deeply 
into the flux density distribution for putative repeating 
FRBs.

In any case, it is possible that with more time on source, some of the Parkes FRBs will
be found to be repeating. Based on current observations, however,
we cannot exclude the possibility that Parkes FRBs represent a
non-repeating population, and thus are fundamentally different from \src.
The differing pulse widths and spectra as discussed above may provide
modest support for this idea.

\section{Summary and Conclusions}
\label{sec:concl}

Our discovery of repeating bursts from \src\ shows that,
for at least one source, the origin of the bursts cannot be cataclysmic, and
further, must be able to repeat on short (\lapp 1\,min) time scales.
Whether \src\ is a unique object in the currently known sample 
of FRBs, or all FRBs are capable of repeating, 
its characterization is extremely important to understanding
fast extragalactic radio transients.

Here, we have shown that bursts from \src\ are detected at 2\,GHz (with GBT)
as well as 1.4\,GHz (with Arecibo). 
The spectra of those bursts are also not well described by
a typical power law and vary significantly from burst to burst.
These variations cannot be due to diffractive interstellar scintillation
and are therefore likely intrinsic. As noted by \citet{ssh+16} the spectral 
variations are somewhat reminiscent of those sometimes seen in the Crab pulsar. 
The episodic burst rate that
we observe from \src\ could be explained by modulation by refractive interstellar
scintillation, but intrinsic explanations based on phenomena displayed by
Galactic pulsars and magnetars could also work.

We have also presented observations of the field from the VLA, 
and the \cxo\ and \swift\ X-ray telescopes, as well as archival 
optical/IR observations from WISE and
IPHAS. None of these observations shows an obvious counterpart to \src. 
Further, from these observations, we have placed a limit on the existence of a 
Galactic nebula that provides the excess dispersion. We find it extremely 
unlikely that \src\ is Galactic, as any nebula that could provide the observed 
dispersion should be visible in VLA, WISE, and/or IPHAS observations.

The nearly certain extragalactic distance and repeating nature of \src\
lead us to favor an origin for the bursts that invokes a young extragalactic neutron star.
Super-giant pulses from young pulsars or magnetars \citep{cw16,pc15}
or radio counterparts to magnetar X-ray bursts \citep{lyu02,pp13,kat15} remain plausible models.
As young neutron stars are expected to be embedded in star-forming regions as 
well as supernova remnants, 
i.e. regions composed of a high amount of dispersing plasma, we expect a large
host contribution to the excess DM. This could lead to a smaller distance for
\src\ than the $\sim 1$\,Gpc implied if the majority of the dispersion comes
from the IGM \citep{sch+14}. This would reduce the seemingly extreme 
luminosities required for these distant bursts.

However, the distance will remain uncertain until a host galaxy for \src\
can be identified. Such an identification could occur either by detecting
radio bursts with interferometry in order to achieve the $\sim1\arcsec$
localization required, or by finding correlated variability at other
wavelengths (e.g. coincident X-ray bursts). Such efforts are currently
underway.

Finally, we have updated the observed FRB rate for the PALFA survey given
the longer baseline of observations since that reported by \citet{sch+14} and
find that we can still neither confirm nor refute the claimed Galactic
latitude dependence of the observed FRB rate.  However, our revised rate is now in
disagreement with the expectations of the diffractive interstellar scintillation 
model of \citet{mj15} that
attempts to account for the latitude dependence.  
\\

We thank A.~Lyne for coordinating the Lovell telescope observations and 
P.~Demorest for help with the VLA observations.

P.S. acknowledges support of a Schulich Graduate Fellowship from McGill University.
L.G.S., J.W.T.H, C.G.B., and J.v.L. gratefully acknowledge support from the European 
Research Council under the European Union’s Seventh Framework Programme 
(FP/2007-2013)/ERC Grant Agreement nos.~279702 (L.G.S.), 337062 (J.W.T.H, C.G.B.), and 617199 (J.v.L.)
J.W.T.H. is an NWO Vidi Fellow.
V.M.K. holds the Lorne Trottier and a Canada Research Chair and receives support 
from an NSERC Discovery Grant and Accelerator Supplement, from a R. Howard 
Webster Foundation Fellowship from the Canadian Institute for Advanced Research (CIFAR), 
and from the FRQNT Centre de Recherche en Astrophysique du Quebec. 
J.S.D. was supported by the NASA Fermi Guest Investigator program and the Chief of Naval Research.
Pulsar research at UBC is supported by an NSERC Discovery Grant and by CIFAR.

We thank the staff of the Arecibo Observatory, and in particular A. Venkataraman, H. Hernandez, P. Perillat and J. Schmelz, for
their continued support and dedication to enabling observations like those presented here.

The Arecibo Observatory is operated by SRI International under a cooperative agreement with the National Science Foundation (AST-1100968), 
and in alliance with Ana G. M\'endez-Universidad Metropolitana, and the Universities Space Research Association.

The National Radio Astronomy Observatory is a facility of the National Science Foundation operated under cooperative agreement by Associated Universities, Inc.

This work is partially based on observations with the 100-m telescope of the MPIfR
(MaxPlanck-Institut f\"{u}r Radioastronomie) at Effelsberg.

These data were processed using the McGill University High Performance Computing Centre operated by Compute Canada and Calcul Quebec.

The scientific results reported in this article are based in part on observations made by the \chandra\ X-ray Observatory.

This publication makes use of data products from the Wide-field Infrared Survey Explorer, which is a joint project of the University of California, Los Angeles, and the Jet Propulsion Laboratory/California Institute of Technology, funded by the National Aeronautics and Space Administration.

This paper makes use of data obtained as part of the INT Photometric H$\alpha$ Survey of the Northern Galactic Plane (IPHAS) carried out at the Isaac Newton Telescope (INT). The INT is operated on the island of La Palma by the Isaac Newton Group in the Spanish Observatorio del Roque de los Muchachos of the Instituto de Astrofisica de Canarias. All IPHAS data are processed by the Cambridge Astronomical Survey Unit, at the Institute of Astronomy in Cambridge. The band-merged DR2 catalogue was assembled at the Centre for Astrophysics Research, University of Hertfordshire, supported by STFC grant ST/J001333/1.

This research has made use of the NASA/IPAC Infrared Science Archive, which is operated by the Jet Propulsion Laboratory, California Institute of Technology, under contract with the National Aeronautics and Space Administration.

\bibliographystyle{apj}
\bibliography{journals_apj,myrefs,modrefs_mcgill,modrefs,psrrefs,crossrefs}

\end{document}